\documentclass{WileyMSP-template}

\usepackage{color,MnSymbol}
\newcommand{\GVS}{GaV$_4$S$_8$}
\newcommand{\GVSe}{GaV$_4$Se$_8$}
\newcommand{\GMS}{GaMo$_4$S$_8$}
\newcommand{\GNS}{GaNb$_4$S$_8$}
\newcommand{\GeVS}{GeV$_4$S$_8$}
\newcommand{\TJT}{$T_\mathrm{JT}$}

\begin{document}

\pagestyle{fancy}
\rhead{\includegraphics[width=2.5cm]{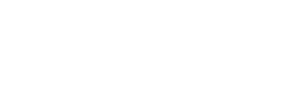}}

\title{Optical, dielectric, and magnetoelectric properties of ferroelectric and antiferroelectric lacunar spinels}

\maketitle


\author{Korbinian Geirhos*}
\author{Stephan Reschke}
\author{Somnath Ghara}
\author{Stephan Krohns}
\author{Peter Lunkenheimer}
\author{Istv\'an K\'ezsm\'arki}



\begin{affiliations}
K. Geirhos, Dr. S. Reschke, Dr. S. Ghara, Dr. S. Krohns, Dr. P. Lunkenheimer, Prof. Dr. I.  K\'ezsm\'arki \\
Experimental Physics V, Center for Electronic Correlations and Magnetism, University of Augsburg, D-86159 Augsburg, Germany\\
E-mail: korbinian.geirhos@physik.uni-augsburg.de
\end{affiliations}


\keywords{Optical Properties, Dielectric Properties, Multiferroics, Magnetoelectric Coupling, Lacunar Spinels}

\begin{abstract}
Lacunar spinels with a chemical formula of $AM_4X_8$ form a populous family of narrow-gap semiconductors, which offer a fertile ground to explore correlation and quantum phenomena, including transition between Mott and spin-orbit insulator states, ferro/ antiferroelectricity driven by cluster Jahn-Teller effect, and magnetoelectric response of magnetic skyrmions with polar dressing. The electronic and magnetic properties of lacunar spinels are determined to a large extent by their molecular-crystal-like structure. The interplay of electronic correlations with spin-orbit and vibronic couplings leads to a complex electronic structure already on the single-cluster level, which -- together with weaker inter-cluster interactions -- gives rise to a plethora of unconventional correlated states. This review primarily focuses on recent progresses in the field  of optical, dielectric, and magnetoelectric properties on lacunar spinels. After introducing the main structural aspects, lattice dynamics and electronic structure of these compounds are discussed on the basis of optical spectroscopy measurements. Dielectric and polarization studies reveal the main characteristics of their low-temperature ferro- or antiferroelectric phases as well as orbital fluctuations in their high-temperature cubic state. Strong couplings between spin, lattice, and orbital degrees of freedom are manifested in singlet formation upon magnetostructural transitions, the emergence of various multiferroic phases, and exotic domain-wall functionalities.   
\end{abstract}


\section{Introduction} \label{intro}

Cubic spinels with the chemical formula $AM_2X_4$ are unarguably one of the largest compound families, comprising a plethora of transition metal ($M$) oxides and chalcogenides \cite{zhao2017, grimes1975, athinarayanan2021}. Lacunar spinels, being a populous sub-class of spinels, have been attracting increasing attention, since they offer a fertile ground for intriguing correlated and quantum states, such as a topological superconductor state \cite{park2020}, spin-orbit entangled molecular $J_\mathrm{eff}$=3/2 states \cite{jeong2017, kim2014}, N\'eel-type skyrmions with polar dressing \cite{ruff2015}, molecular cluster orbital-driven ferroelectricity \cite{ruff2015, singh2014,  fujima2017, xu2015, neuber2018, geirhos2018} and antiferroelectricity \cite{geirhos2020b}, multiple multiferroic phases \cite{ruff2015, fujima2017, geirhos2020, felea2020}, and magnetic states confined to polar domain walls \cite{geirhos2020}. Pressure-induced superconductivity \cite{abd2004}, avalanche-like breakdown of the Mott gap, leading to a resistive switching \cite{guiot2013}, and colossal negative magnetoresistance \cite{dorolti2010, janod2015} further witness the complexity of these compounds.     

\begin{figure}[h]
\centering
  \includegraphics[width=0.9\linewidth]{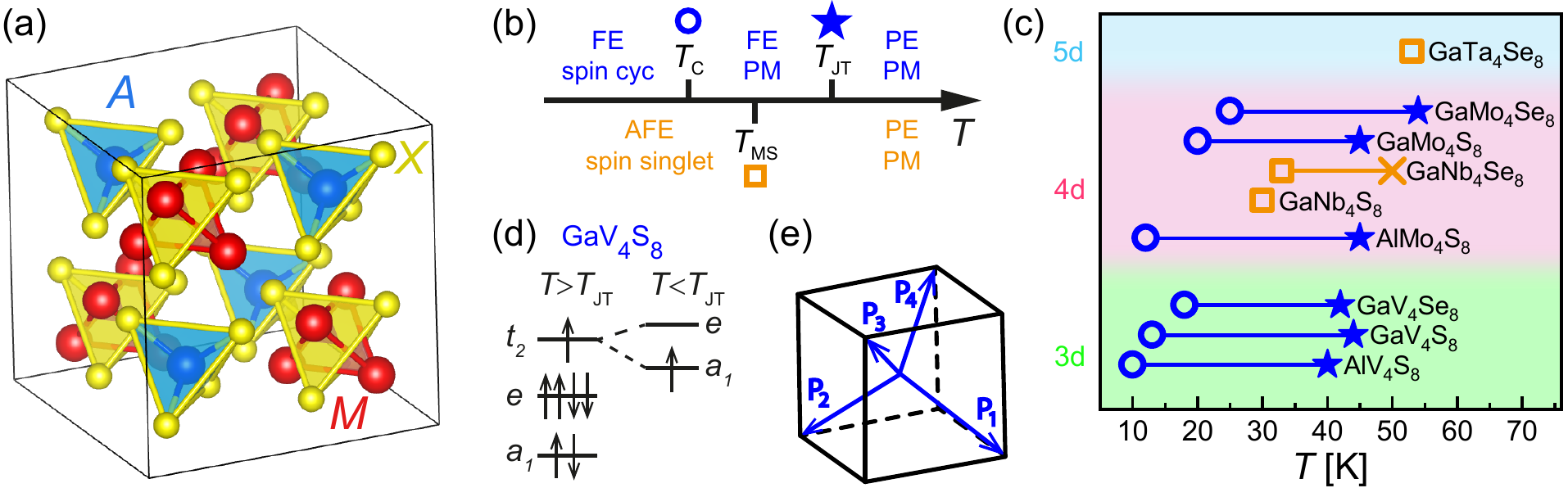}
  \caption{a) Cubic lacunar spinel strucutre at room-temperature. b) Transition paths of spin 1/2 lacunar spinels showing distinct polar and magentic order (PE: paraelectric, PM: paramegentic). c) Transition temperatures of lacunar spinels with different transition metal clusters. Stars denote a structural and ferroelectric transition, circles a magnetic transition into a magnetic modulated phase, and rectangles a magnetostructural transition. The cross for GaNb$_4$Se$_8$ indicates a purely structural transition. d) Energy level scheme of the molecular orbital of the V$_4$S$_4$ unit of \GVS . e) Four polar directions ($P_{1-4}$) of rhombohedral lacunar spinels with respect to the cubic unit cell.}
  \label{Fig_TransTemp}
\end{figure}

The term {\it lacunar} refers to an important structural motif, namely the {\it lack} of every second $A$-site ion with respect to the normal spinel structure, as described by the chemical formula  $A\medsquare M_4X_8$ or simply $AM_4X_8$. 
The regular alternation of voids ($\medsquare$) and cations on the diamond lattice formed by the $A$ sites has two important consequences: i) the inversion symmetry present in the cubic $Fd\bar{3}m$ spinel structure is lost and the symmetry is reduced to the non-centrosymmetric cubic $F\bar{4}3m$ and ii) the pyrochlore lattice of the $M$ sites develops a breathing, where smaller and larger corner-sharing $M_4$ tetrahedra alternate in a regular fashion. The breathing can be such strong that the $M$--$M$ distance in the smaller tetrahedra is close to the interatomic distance of the corresponding elemental metal, as is the case for $M$=V. Correspondingly, the lacunar spinel structure can be described as two interpenetrating fcc networks. One network is formed of weakly linked $M_4X_4$ molecular units composed of interpenetrating $M_4$ and $X_4$ tetrahedra and the other consits of $AX_4$ units, as displayed in \textbf{Figure \ref{Fig_TransTemp}}a. The weak electronic overlap between the $M_4X_4$ cubane units renders these compounds narrow-gap semiconductors \cite{reschke2020}. 

The lacunar spinel structure is realized by a remarkably large number of compounds, including $3d$ (V, Ti), $4d$ (Mo, Nb), and $5d$ (Ta) transition metals in combination with $A$=Ga, Ge, Al, and $X$=S, Se, Te. Due to their molecular-crystal-like structure, the orbital scheme of the $M_4$ tetrahedra is often considered as a good starting point to describe the electronic structure of lacunar spinels \cite{johrendt1998, pocha2000, pocha2005, jakob2007, hozoi2020}. In fact, the tendency to form cluster orbitals is widely accepted for the V- and Mo-based compounds, like GaV$_4$S(e)$_8$ and GaMo$_4$S(e)$_8$ \cite{johrendt1998, pocha2000, bichler2010}, where the uppermost partially occupied molecular orbital is a triply degenerate $t_2$ level, hosting a single unpaired electron and hole, respectively (see Figure~\ref{Fig_TransTemp}d).

This orbital degeneracy is lifted by a cooperative Jahn-Teller distortion at \TJT, which reduces the symmetry to polar rhombohedral ($R3m$). Upon this ferroelectric/ferroelastic transition, all $M_4X_4$ units within a structural domain get distorted, elongated in GaV$_4$S(e)$_8$ and compressed in GaMo$_4$S(e)$_8$,along the same vertex of the $M_4$ tetrahedra (i.e., one of the cubic $<$111$>$-type axes), which also lowers their point-group symmetry to $3m$ \cite{pocha2000, bichler2010}. Due to the lack of inversion in the parent cubic state, only four polar domain states can form \cite{neuber2018, kezsmarki2015, butykai2017, bordacs2017}. Their polarizations ($P$) span 109$^{\circ}$ with each other, i.e., there are no $\pm P$ domain states coexisting (see Figure~\ref{Fig_TransTemp}e). These polar domains have been directly imaged by piezoelectric force microscopy in GaV$_4$S$_8$ \cite{butykai2017}, GaV$_4$Se$_8$ \cite{geirhos2020} and GaMo$_4$S$_8$ \cite{neuber2018}. The ferroelectric transitions can be categorized as order-disorder-type, since the existence of local electric dipoles above \TJT\ and the dynamic Jahn-Teller effect associated with their fluctuations have been evidenced in GaV$_4$S$_8$ by dielectric studies \cite{geirhos2018, wang2015, ruff2017, reschke2017}. 

In addition to the onset of ferroelectricity (FE), the rhombohedral distortion also affects the magnetic interactions. The overall exchange coupling, originally being antiferromagnetic in the cubic state, turns to ferromagnetic below \TJT\ \cite{ruff2017,widmann2017, powell2007}. Moreover, the axial polar distortion generates a Dzyaloshinskii-Moriya interaction pattern that can stabilize cycloidal (cyc) modulations with propagation vectors perpendicular to the polar axis. In fact, the emergence of cycloidal spin order has been observed in these compounds below $T_\mathrm{C}$. In finite magnetic fields the cycloidal order can be transformed to a N\'eel-type skyrmion lattice (SkL) state and subsequently to a field-polarized ferromagnetic (FM) state \cite{kezsmarki2015, bordacs2017, butykai2019}. Importantly, the ferroelectric and the magnetic transitions are decoupled from each other as shown in Figure~\ref{Fig_TransTemp}b and c. The full list of transition temperatures are given in \textbf{Table~\ref{Table_Temperatures}}.

\begin{table}[h]
\centering
\caption{\label{Table_Temperatures} Jahn-Teller \TJT , magnetic $T_\mathrm{C}$, and magnetostructural $T_\mathrm{MS}$ transition temperatures of lacunar spinels. The temperature marked by the asterisk most likely is a simple structural transition, which is not Jahn-Teller driven \cite{ishikawa2020}.}

\begin{tabular}[htbp]{ @{}lllll@{}}
\hline
 & \TJT\,(K) &  $T_\mathrm{C}$\,(K) & $T_\mathrm{MS}$\,(K) & Low-T structure \\
 \hline
 AlV$_4$S$_8$ & 40 \cite{bichler2010} & 10 \cite{bichler2010} & & $R3m$ \cite{bichler2010}  \\
 \GVS & 44 \cite{widmann2017} & 13 \cite{widmann2017} & & $R3m$ \cite{pocha2000} \\
 \GVSe & 41 \cite{ruff2017} & 18 \cite{ruff2017} & & $R3m$ \cite{bichler2010} \\
 AlMo$_4$S$_8$ & 45 \cite{francois1992} & 12 \cite{francois1992} & & $R3m$ \cite{francois1992} \\
 \GNS & & & 31 \cite{jakob2007} & $P2_12_12_1$ \cite{geirhos2020b} \\ 
 GaNb$_4$Se$_8$ & 50* \cite{ishikawa2020} & & 33 \cite{ishikawa2020}& $P2_12_12_1$ \cite{langmann2021}  \\
 \GMS & 45 \cite{pocha2000} & 20 \cite{pocha2000} & & $R3m$ \cite{pocha2000} \\
 GaMo$_4$Se$_8$ & 54 \cite{rastogi1983} & 25 \cite{rastogi1983} & & $R3m$ \cite{francois1992}  \\
 GaTa$_4$Se$_8$ &  &  & 53 \cite{ishikawa2020} & $P2_12_12_1$ \cite{langmann2021} \\
  GeV$_4$S$_8$ & 30 \cite{widmann2016}  &  & 15 \cite{widmann2016} & $Imm2$ \cite{muller2006} \\
  \hline
\end{tabular}

\end{table}

As discussed above, the Jahn-Teller activity of the $M_4$ cluster leads to the onset of orbital-order-driven ferroelectricity below \TJT\ in GaV$_4$S(e)$_8$ and GaMo$_4$S(e)$_8$, followed by magnetic ordering at $T_\mathrm{C}<$ \TJT.
However, although the $M_4$ clusters have the same formal valency in GaNb$_4$S(e)$_8$ and GaTa$_4$S(e)$_8$ as in the compounds above and the main axis of distoriton of the $M_4$ tetrahedra is also along one of the vertices, the symmetry lowering in these cases is realized via a single magnetostructural transition at $T_\mathrm{MS}$, which lowers the cubic symmetry to orthorhombic (see Figure~\ref{Fig_TransTemp}b and c) \cite{geirhos2020b,ishikawa2020, langmann2021}. 
Specific to GaNb$_4$Se$_8$, at first the face centering is lost by a transition from the $F\bar{4}3m$ parent state to the cubic chiral $P2_13$ state, followed by the loss of the cubic symmetry at $T_\mathrm{MS}$.

In terms of the polar degrees of freedom, the low-temperature orthorhombic phase ($T<T_\mathrm{MS}$) is antiferroelectric (AFE) \cite{geirhos2020b}. Antiferroelectricity is realized by the non-zero polarization of individual distorted $M_4X_4$ clusters, ordering in an alternating fashion. The magnetostructural nature of the transition is best revealed in GaNb$_4$S$_8$, where the quenching of spin degrees of freedom via spin-singlet formation at $T_\mathrm{MS}$ was evidenced by NMR spectroscopy \cite{jakob2007,waki2010}. Similar to the materials with ferroelectric ground state, the fluctuation of local polar distortions has been observed already above $T_\mathrm{MS}$, i.e., in the nominally cubic state \cite{geirhos2020b, waki2010}.

We want to point out, that the first lacunar spinel which was shown to become ferroelectric upon the structural transition was GeV$_4$S$_8$ \cite{singh2014}. However, in contrast to all aforementioned lacunar spinels, the highest molecular orbital level of the $M_4$ cluster in this compound is occupied by two unpaired electrons with total spin $S=1$. This leads to a structural change to the polar orthorhombic space group $Imm2$ upon \TJT\ and an elongation and shortening of two opposite V-V bonds of the V$_4$ tetrahedra, respectively. Moreover, this compound also shows an additional strucural change upon its antiferrmoagentic ordering \cite{johrendt1998, muller2006, bichler2008}

The intimate coupling between spin, orbital, and lattice degrees of freedom together with the competition/interplay of various types of interactions lead to the aforementioned fascinating magnetic and electric properties in lacunar spinels but also make their microscopic description challenging. From the experimental side, this challenge can be addressed, in a first step, by studying the electronic structure and the lattice dynamics of these compounds by optical spectroscopy. The optical properties of lacunar spinels are summarized in section~\ref{optic}. Systematic optical spectroscopy studies are of particular importance, because of fundamental difficulties in modelling the electronic structure of lacunar spinels, requiring theoretical tools beyond standard density-functional theory calculations, such as cluster dynamical mean-field theory \cite{kim2020} and quantum chemistry approaches \cite{hozoi2020}. Splittings of optical phonon modes provide a sensitive probe of the symmetry lowering upon \TJT\ and $T_\mathrm{MS}$ \cite{reschke2020, hlinka2016}, while broadband optical spectroscopy can follow reconstructions of the gap edge associated with these structural transitions \cite{reschke2020}.

As a next step, the cooperative behavior of polar constituents can be revealed by dielectric spectroscopy and domain imaging techniques. In particular, polarization measurements and dielectric spectroscopy provide valuable information about the nature of the polar state. Broadband dielectric spectroscopy over the Hz--THz range also enables the investigation of intriguing dipolar dynamics around the non-canonical polar or antipolar transitions of these materials, which are discussed in section~\ref{diel}. Studies on the ferro-, but also the magnetoelectric polarization are summarized in section~4. The investigation of the magnetoelectric phenomena can give insight into the coupling of orbital and lattice degrees of freedom to spins. The present review summarizes the progress achieved along these lines and discusses open questions and future directions.

\section{Optical properties} \label{optic}

For theory it still remains a challenge to describe the electronic structure of the lacunar spinels. Based on a simple electron counting, one would expect metallic behavior due to the partial occupation of $t_2$ orbitals of $M_4X_4$ clusters. However, lacunar spinels are found to be semiconductors already in their high-temperature cubic state. Standard density-functional theory calculations on the DFT+$U$ level \cite{johrendt1998,pocha2005,shanthi1999,camjayi2012,cannuccia2017,kim2018,wang2019,zhang2020} have failed to reproduce the finite gap of the cubic phase observed experimentally \cite{guiot2013,reschke2020}. Difficulties in the modelling of the electronic structure of the lacunar spinels arise from the interplay of various factors. The character of the $M_4X_4$ cluster orbitals, which governs the optical, dielectric, and magnetic properties, is determined by a delicate balance between the Jahn-Teller instability of the clusters, vibronic interactions as well as spin-orbit and spin-lattice couplings. Moreover, the character of the cluster orbitals strongly affects the inter-cluster hopping. Therefore, reproducing the narrow-gap semiconductor nature of these compounds may require advanced theoretical approaches going beyond the usual density-functional theory schemes.

\begin{figure}[h]
  \centering
  \includegraphics[width=0.5\linewidth]{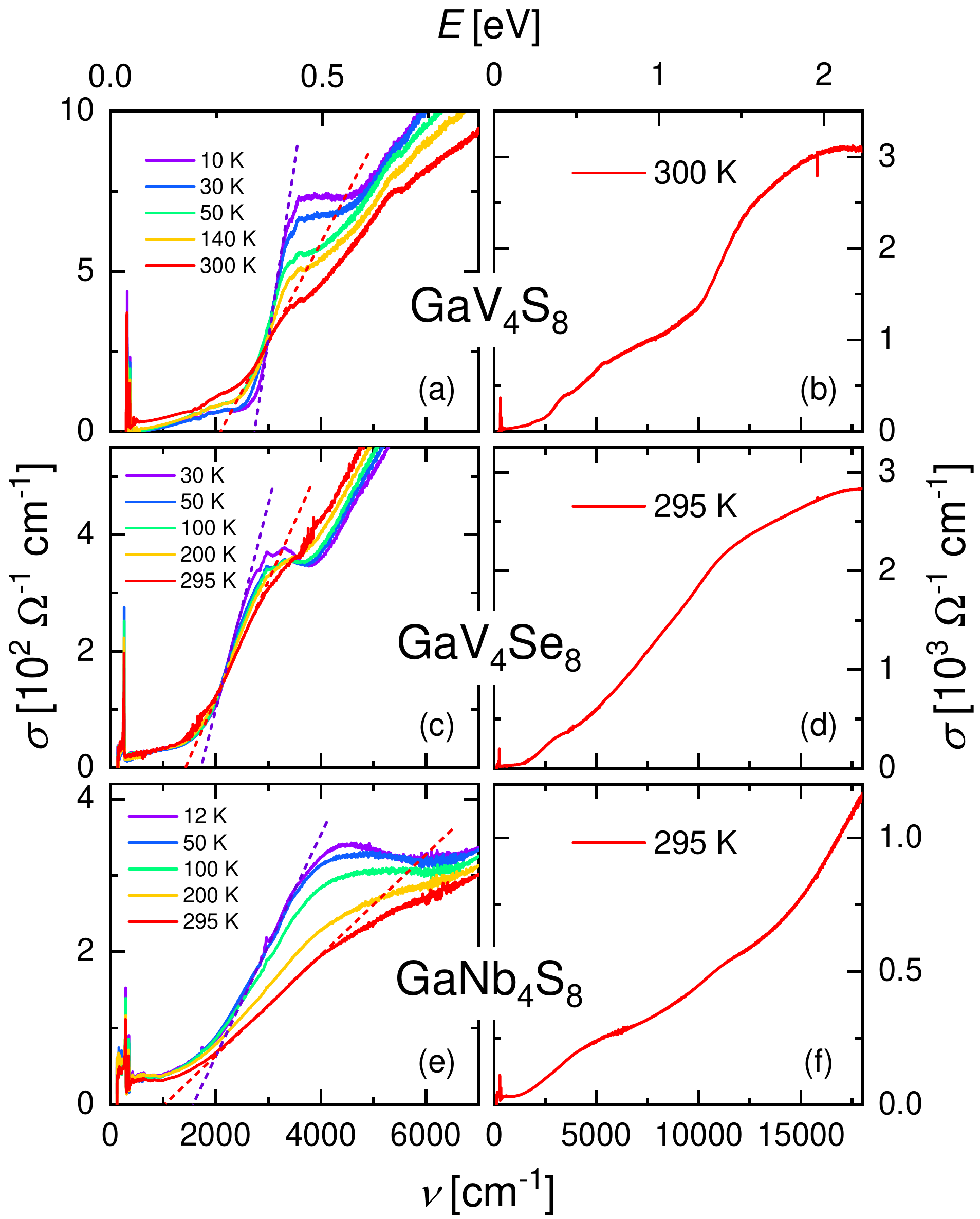}
  \caption{Frequency-dependent optical conductivity of a) and b) GaV$_4$S$_8$, c) and d) GaV$_4$Se$_8$ and e) and f) GaNb$_4$S$_8$ for selected temperatures in the region of the band gap (left panels) \cite{reschke2020}. Dashed lines indicate a linear extrapolation of the increase of the conductivity for determination of the gap energy. Corresponding broadband optical conductivity at room temperature (right panels). Adapted with permission.\textsuperscript{\cite{reschke2020}} Copyright 2020, American Physical Society.}
  \label{Fig_OptCond}
\end{figure}

Broadband optical spectroscopy can provide a deep insight into the electronic band structure, especially when a series of compounds is systematically investigated, by probing the coupled density of states for electric dipole-active transitions. \textbf{Figure \ref{Fig_OptCond}}b, d, and f show the room temperature broadband optical conductivity of GaV$_4$S$_8$, GaV$_4$Se$_8$, and GaNb$_4$S$_8$  up to 18000\,cm$^{-1}\approx2.25$\,eV, respectively,
as obtained from reflectivity spectra via Kramers-Kronig transformation \cite{reschke2020, reschke2017}.
An enlarged view of the low-energy optical conductivity (Figure \ref{Fig_OptCond}a,c,e) reveals the temperature-induced evolution of electronic transitions located below 7000\,cm$^{-1}\approx0.88$\,eV. For all temperatures in the range of 10 - 300\,K the optical conductivity vanishes in the zero-frequency limit, which confirms the semiconducting nature of these lacunar spinels, already in the high-temperature cubic phase. In the spectral range between 2000 - 4000\,cm$^{-1}$ the optical conductivity of these compounds exhibits a sudden increase, associated with the electronic band gap. The gap edge is manifested by a pronounced and sudden increase of the conductivity at low temperatures, which is continuously smeared out with increasing temperature. This temperature-dependent behavior has primarily been ascribed to orbital fluctuations in the high-temperature cubic phase \cite{reschke2020,reschke2017}. In fact, the dynamic Jahn-Teller effect, i.e. the dynamical breaking of cubic symmetry by local distortions, is a possible reason for the failure of DFT+$U$ calculations to reproduce the narrow but finite gap in the high-temperature phase.

The size of the band gap has been determined by a linear extrapolation of the optical conductivity to zero, as illustrated by dashed lines in Figure \ref{Fig_OptCond}a,c,e. At room temperature, this procedure yields the following gap energies: 260\,meV for GaV$_4$S$_8$, 175\,meV for GaV$_4$Se$_8$, and 130\,meV for GaNb$_4$S$_8$ \cite{reschke2020,reschke2017}. These gap energies are in reasonable agreement with values previously reported based on resistivity measurements \cite{pocha2005, widmann2017,bichler2011}, characterizing the lacunar spinels as narrow-gap semiconductors. A systematic study, involving lacunar spinels with $3d$, $4d$, and $5d$ transition metal clusters \cite{reschke2020}, reveals that the charge gap is reduced from $3d$ to $5d$ compounds. With decreasing temperature the band gap increases gradually in all investigated compounds with tiny anomalies of the gap energies at the structural transitions \cite{reschke2020}. Though the values of the charge gap are not strongly temperature dependent, a considerable temperature-induced spectral weight transfer is observed in each compound near the gap edge, which may originate from the slowing down of the orbital fluctuations with decreasing temperature and the onset of static cooperative order upon the structural transition \cite{reschke2020,reschke2017}. In this scenario, vibronic interactions should be essential ingredients in describing the electronic structure of lacunar spinels. While these optical studies quantify the charge gap and its variation from $3d$ to $5d$ compounds, the determination of the electronic structure in the vicinity of the Fermi energy calls for theory support. A recent resonant inelastic X-ray scattering study implies that Ta$_4$ clusters in the $5d$ GaTa$_4$Se$_8$ may offer a realization of molecular $J_\mathrm{eff}$=3/2 states \cite{jeong2017}, indicating strong spin-orbit effects. In contrast, the structural transitions of several $3d$ and $4d$ lacunar spinels are well described by the simple orbital Jahn-Teller effect, implying a secondary role of spin-orbit coupling in these compounds.  

\begin{figure}[h]
  \centering
  \includegraphics[width=0.5\linewidth]{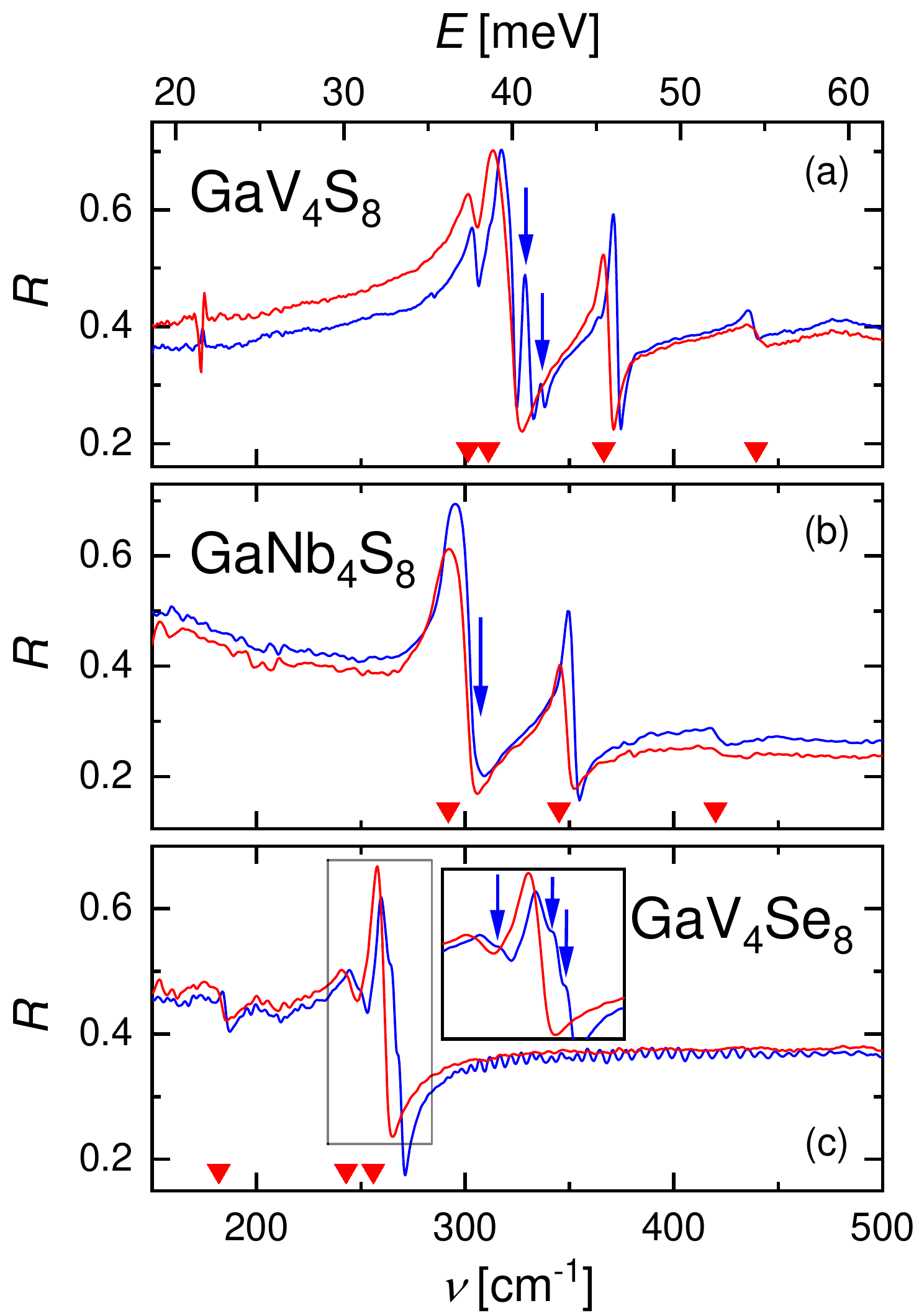}
  \caption{Frequency-dependent phonon reflectivity spectra of a) GaV$_4$S$_8$, b) GaNb$_4$S$_8$, and c) GaV$_4$Se$_8$ measured at room temperature (red curves) and at 10\,K (blue curves) \cite{reschke2020}. Red triangles indicate the mode frequencies at room temperature. Blue arrows indicate additional phonon modes appearing only below the structural transition temperature. Adapted with permission.\textsuperscript{\cite{reschke2020}} Copyright 2020, American Physical Society.}
  \label{Fig_Phonons}
\end{figure}

The study of the infrared (IR) active phonon modes is often a highly sensitive tool to investigate changes of the crystal structure at the phase transitions \cite{bordacs2009,kant2010,kocsis2013}. Generally, the number of the IR-active optical phonon modes can be determined by group theory based on the crystal symmetry and the occupied Wyckoff positions. For the high-temperature cubic phase of the lacunar spinels with space group $F\overline{4}3m$, six triply degenerate IR-active $F_2$ phonon modes are expected.
Below \TJT{} and $T_\mathrm{MS}$, the symmetry of the studied compounds is lowered to rhombohedral (\GVS, \GVSe) with space group $R3m$ \cite{pocha2000} and orthorhombic with space group $P2_12_12_1$ (\GNS) \cite{geirhos2020b}, respectively. These symmetry lowering transitions should increase the number of IR-active zone center phonon modes to 21 and 141, respectively.

Far-IR reflectivity spectra of GaV$_4$S$_8$, GaV$_4$Se$_8$, and GaNb$_4$S$_8$ measured at room temperature and at 10\,K are shown in \textbf{Figure~\ref{Fig_Phonons}} \cite{reschke2020}. These spectra reveal only a subset of the six IR-active phonon modes expected in the high-temperature cubic phase: Four modes are visible in GaV$_4$S$_8$ (a) and three modes in GaNb$_4$S$_8$ (b) and GaV$_4$Se$_8$ (c), as seen in Figure \ref{Fig_Phonons}. The remaining modes obviously have too low spectral weight to be detected by reflectivity measurements. 

The room temperature mode frequencies, as indicated by the red triangles in Figure~\ref{Fig_Phonons}, are very similar for GaV$_4$S$_8$ (a) and GaNb$_4$S$_8$ (b). In contrast, the phonon eigenfrequencies of GaV$_4$Se$_8$ (c) are considerably lower, which can be explained by the mass ratio of S and Se. A systematic phonon study over a wide series of lacunar spinels, where $A$, $M$ and, $X$ atoms were sequentially changed, revealed that the modes observed in the high-temperature cubic phase can be assigned to vibrations of the ligands and the $A$-site ions \cite{reschke2020}. In other words, this implies that the vibrations of the tetrahedral $M_4$ units carry only a weak electric dipole moment. This is a surprising observation in the context of the widely accepted picture of the cubic-to-rhombohedral transition driven by the polar distortions of the $M_4$ clusters.       

Although in the distorted state of these compounds again a subset of the expected modes are observed, the low-temperature phonon spectra allow to trace the symmetry lowering taking place at the structural transition. As indicated by the blue arrows in Figure~\ref{Fig_Phonons}, additional weak modes appear below $T_\mathrm{JT}$ and $T_\mathrm{MS}$, indicating the cubic-to-rhombohedral and the cubic-to-orthorhombic structural phase transitions, respectively. The low number of additional phonon modes detectable upon the structural transition is in agreement with the small deviations from the cubic structure, observed in these compounds \cite{pocha2000,jakob2007}. Though spin-lattice coupling plays an important role in the studied compounds, no anomalies in the phonon spectra have been detected at $T_\mathrm{C}$.

\section{Dielectric properties}
\label{diel}

The investigation of lacunar spinels by dielectric spectroscopy is mainly motivated by the polar order detected in many of these materials as mentioned in section \ref{intro} \cite{ruff2015, singh2014, xu2015, neuber2018, geirhos2018, geirhos2020b,ruff2017}. In recent years, unconventional mechanisms for the generation of polar order have come into the focus of interest. Such mechanisms often promote multiferroicity, especially the simultaneous ordering of spins and electrical dipoles, being of fundamental interest as well as relevant for applications in future electronics and spintronics \cite{spaldin2005}. An interesting example is the ferroelectric order triggered by orbital ordering in various materials, including several lacunar spinels \cite{ruff2015, singh2014, keimer2006, barone2011, barone2012, yamauchi2014}. As discussed in section \ref{intro}, the Jahn-Teller distortion of the $M_4X_4$ cubane units produces a dipole moment, and the collective ferro-type ordering of these local dipoles generates ferroelectricity. Since these cubane units are weakly linked molecule-like entities, these lacunar spinels show analogy with molecular polar materials \cite{shi2016,zhang2019b}.

In general, at a polar phase transition pronounced anomalies are expected to occur in the temperature-dependent dielectric constant $\varepsilon'$ (the real part of the dielectric permittivity). Therefore, aside of electrical polarization measurements (cf. section \ref{pol}), dielectric spectroscopy is an important tool for detecting polar phase transitions and the dipolar dynamics associated with the polar ordering and it can be used for characterizing the order of the transition. Investigating the polar dynamics allows, e.g., to distinguish between the two classical types of ferroelectricity, which can be of displacive or order-disorder nature \cite{blinc1994,lines1996}. In the first case, the non-polar high-symmetry structure, lacking electric dipoles even on the microscopic scale, transforms into a polar structure, where polarization arises from the displacements of ionic cores and/or the deformations of orbitals. In such materials, significant excitations associated with the dipolar dynamics are expected only in the optical range, e.g., the well-known soft-phonon modes. In contrast, in this class of ferroelectrics, $\varepsilon'$ exhibits no significant frequency dependence within the typical frequency range of conventional dielectric spectroscopy (Hz--GHz) and, thus, the polar transition is manifested by an essentially frequency-independent peak in $\varepsilon'(T)$.

In marked contrast, in order-disorder ferroelectrics, the polar order arises from the ordering of dynamically disordered permanent dipole moments that already exist above the ferroelectric ordering temperature. These dipoles can rearrange within multiwell potentials which leads to the typical spectral signatures of relaxational processes in the dielectric frequency range, both above and below the transition. Within the kinetic Ising model used to describe order-disorder ferroelectrics, a critical slowing down of this dipolar dynamics is predicted when approaching the ferroelectric transition from high or low temperatures \cite{blinc1994,lines1996}. Moreover, the peak in $\varepsilon'(T)$, also showing up at the transition in this class of ferroelectrics, becomes successively suppressed with increasing frequency. 

While all these findings are well documented for conventional ferroelectrics \cite{hatta1968,gesi1970,deguchi1992,staresinic2006}, the dipolar dynamics and dielectric anomalies associated with polar transitions are rarely investigated for ferroelectrics with unconventional ordering mechanisms \cite{lunkenheimer15, krohns19}. It is interesting to check, whether the above-described dielectric features of conventional displacive and order-disorder ferroelectrics can also be observed for such non-canonical polar materials.

\begin{figure}[h]
  \centering
  \includegraphics[width=0.5\linewidth]{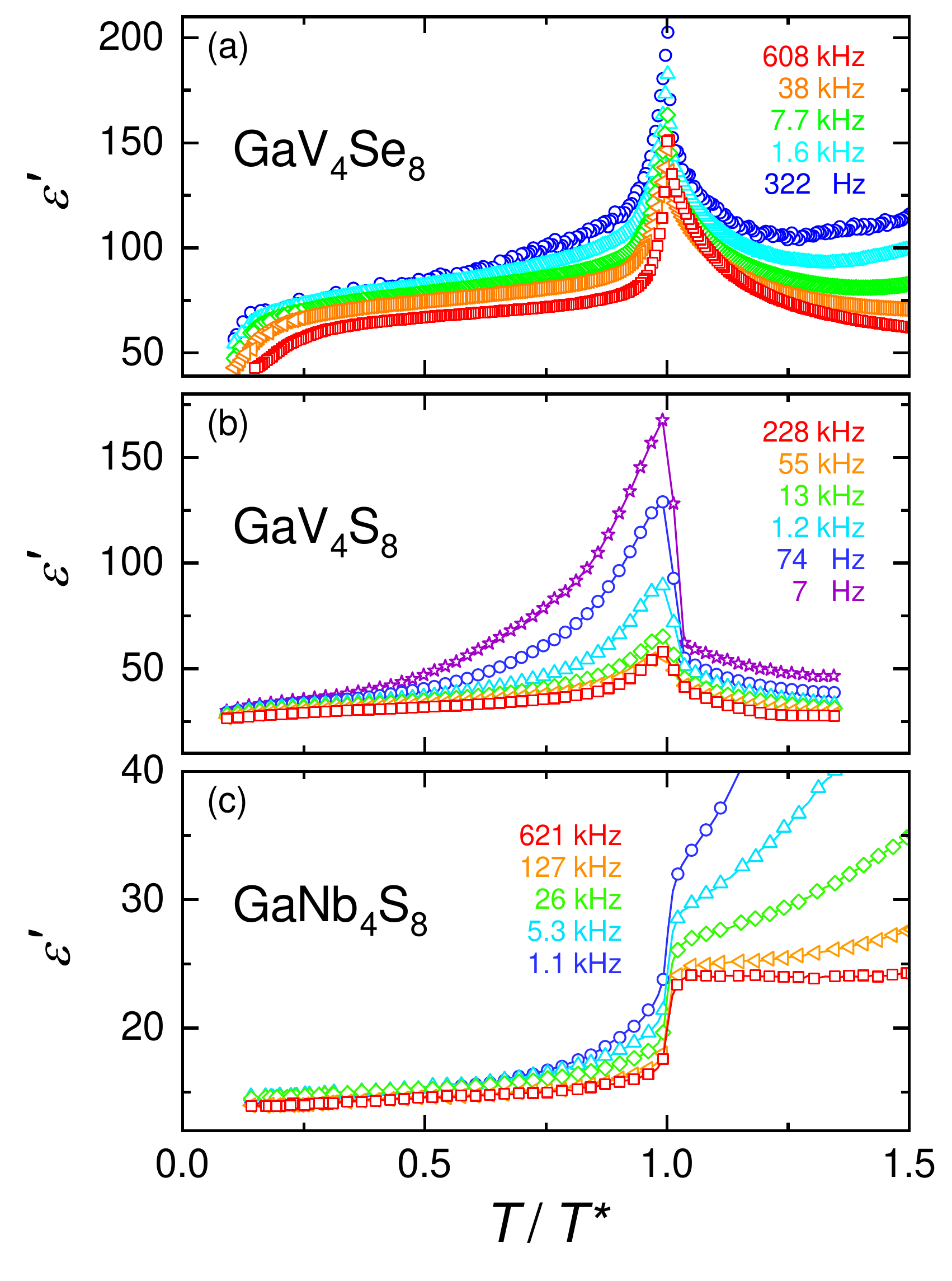}
  \caption{Temperature dependence of the dielectric constant of a) \GVSe\
  \cite{ruff2017}, b) \GVS, and c) \GNS\ \cite{geirhos2020b} for various frequencies with the temperature scale normalized to $T^*$, which represents \TJT\ for \GVSe\ and \GVS\ and $T_\mathrm{MS}$ for \GNS (see Table \ref{Table_Temperatures}). Adapted with permission.\textsuperscript{\cite{ruff2017}} Copyright 2017, American Physical Society. \textsuperscript{\cite{geirhos2020b}} Copyright 2021, American Physical Society.
  }
  \label{Fig_EpsT}
\end{figure}

\textbf{Figure \ref{Fig_EpsT}} shows three examples of the temperature-dependent dielectric constant of polar lacunar spinels, as measured at different frequencies in \GVSe\ (a) \cite{ruff2017}, \GVS\ (b), and \GNS\ (c) \cite{geirhos2020b}. To facilitate the comparison of the three materials, the temperature scale was normalized to \TJT.
For \GVSe\ and \GVS, Figure \ref{Fig_EpsT} reveals qualitatively similar behavior of $\varepsilon'(T)$ while \GNS\ behaves markedly different. As mentioned in section \ref{intro}, in contrast to \GNS, both \GVSe\ and \GVS\ become ferroelectric at the Jahn-Teller transition. Both systems exhibit complex magnetic phase diagrams, including a skyrmion lattice phase \cite{ruff2015, fujima2017, bordacs2017, ruff2017, widmann2017} and, notably, \GVS\  was the first bulk material, where a N\'eel-type skyrmion lattice state was observed \cite{ruff2015, kezsmarki2015}.

In refs. \cite{ruff2015} and \cite{ruff2017}, the orbital-driven polar order of these two compounds\ was evidenced by dielectric and polarization measurements  (cf. section \ref{pol} for the latter). As shown in Figure \ref{Fig_EpsT}a and b, for both systems, $\varepsilon'(T)$ exhibits a clearly pronounced peak at \TJT\ for each frequency. For \GVSe, such a peak was also reported in ref. \cite{fujima2017} for a single frequency. Dielectric anomalies associated with the polar ordering in \GVS\ were also found in ref. \cite{lal2018}. For the latter compound, it was demonstrated \cite{ruff2015, widmann2017} that the peak in $\varepsilon'(T)$ also persists for frequencies up to 2.5~GHz, thus excluding non-intrinsic, electrode-related effects in this region, which play no role at very high frequencies \cite{lunkenheimer15, lunkenheimer2010}. 

Both above and below \TJT, Figure \ref{Fig_EpsT}a and b reveal significant frequency dispersion of $\varepsilon'(T)$ \cite{ruff2015, ruff2017, widmann2017}. 
In particular, the peak amplitude reduces with increasing frequency, pointing to polar order of order-disorder ferroelectric nature \cite{blinc1994,lines1996}. As discussed in detail in refs. \cite{ruff2015, ruff2017}, at high temperatures (beyond the temperature range covered in Figure \ref{Fig_EpsT}a and b) the frequency dispersion strongly increases  and huge values of $\varepsilon'$ are observed, which is due to non-intrinsic contributions. This is known to often arise from electrode effects in semiconducting samples such as the lacunar spinels and special care has to be taken to avoid misinterpretations in terms of intrinsic relaxation processes \cite{lunkenheimer15, lunkenheimer2010}. At sufficiently low temperatures and especially below \TJT, intrinsic relaxational behavior, again typical for order-disorder ferroelectrics \cite{lines1996}, leads to the observed dispersion \cite{ruff2015,ruff2017}. 
In \GVS, pronounced relaxational dispersion effects were observed up to the THz range \cite{wang2015}, which will be discussed in more detail below.

For \GVSe, at the lowest temperatures, the classical signature of canonical relaxation processes in $\varepsilon'(T)$ \cite{lunkenheimer15,kremer2002} shows up: A step-like decrease, which shifts to lower temperatures with decreasing frequency. The latter reflects the essentially thermally activated behavior of conventional relaxation dynamics \cite{lunkenheimer15, kremer2002}, in contrast to the critical temperature dependence expected just below the polar transition \cite{blinc1994,lines1996}. In ref. \cite{ruff2017}, it was speculated that this relaxation could be due to a decoupling of orbital and lattice dynamics, which is affected by the onset of magnetic order. In contrast, the dielectric constant of \GVSe\ does not show any clear anomaly or relaxation step at the magnetic transition.

Overall, with a peak in $\varepsilon'(T)$ that becomes suppressed with increasing frequency and significant relaxation dynamics, both \GVSe\ and \GVS\ reveal signatures of order-disorder ferroelectricity, where fluctuating dipoles already exist above the ferroelectric transition. As the polar and orbital dynamics in these materials are highly entangled, this also implies orbital fluctuations above \TJT, i.e., a dynamic Jahn-Teller effect, in accord with Raman- and IR-spectroscopic studies \cite{reschke2020, hlinka2016} (cf. section \ref{optic}). As corroborated by polarization experiments (section \ref{pol}), these compounds are type-I multiferroics, where the magnetic and ferroelectric order develop independently.

As discussed in section \ref{intro}, in \GVSe\ and \GVS\ the dipole moments are essentially generated by the elongation of the $M_4$ clusters (with $M=$~V for these systems) along the crystallographic $\langle111\rangle$ direction. This is also the case for \GNS\, whose temperature-dependent dielectric constant is shown in Figure \ref{Fig_EpsT}c \cite{geirhos2020b}. However, while in \GVSe\ and \GVS\ all $M_4$ clusters become distorted along the same cubic body diagonal, leading to parallel orientations of the dipole moments and macroscopic polarization, this is not the case for \GNS\ \cite{jakob2007}. Instead, here below $T_\mathrm{MS}$ the distortions arise along four different cubic body diagonals, leading to a non-collinear polar order with zero net polarization. As mentioned in section  \ref{intro}, the resulting structure is compatible with antiferroelectricity, for which markedly different dielectric behavior compared to ferroelectrics is predicted \cite{toledano2016, toledano2019}. Indeed, in contrast to \GVSe\ and \GVS, $\varepsilon'(T)$ of this material (Figure \ref{Fig_EpsT}c) does not show a peak at \TJT, but instead reveals an abrupt reduction below \TJT. This is in accord with the predictions \cite{toledano2016, toledano2019} and with findings for classical antiferroelectrics like KCN or NH$_4$H$_2$PO$_4$ \cite{gesi1972,mason1952}.  

In  Figure \ref{Fig_EpsT}c, significant frequency dispersion of $\varepsilon'$ also shows up for \GNS, both above and below the transition. For elevated temperatures, beyond the region covered in Figure \ref{Fig_EpsT}c, again huge values of $\varepsilon'$ are reached pointing to non-intrinsic effects \cite{lunkenheimer15, lunkenheimer2010}. However, in vicinity and below $T_\mathrm{MS}$, intrinsic dipolar fluctuations seem to dominate. Again, due to the coupling of dipolar and orbital dynamics, this indicates a dynamic Jahn-Teller effect in this material, just as for \GVSe\ and \GVS. This is in accord with the finding of structural fluctuations even far above $T_\mathrm{MS}$, based on nuclear magnetic resonance and X-ray diffraction experiments \cite{geirhos2020b, waki2010}. 

\begin{figure}[h]
  \centering
  \includegraphics[width=0.5\linewidth]{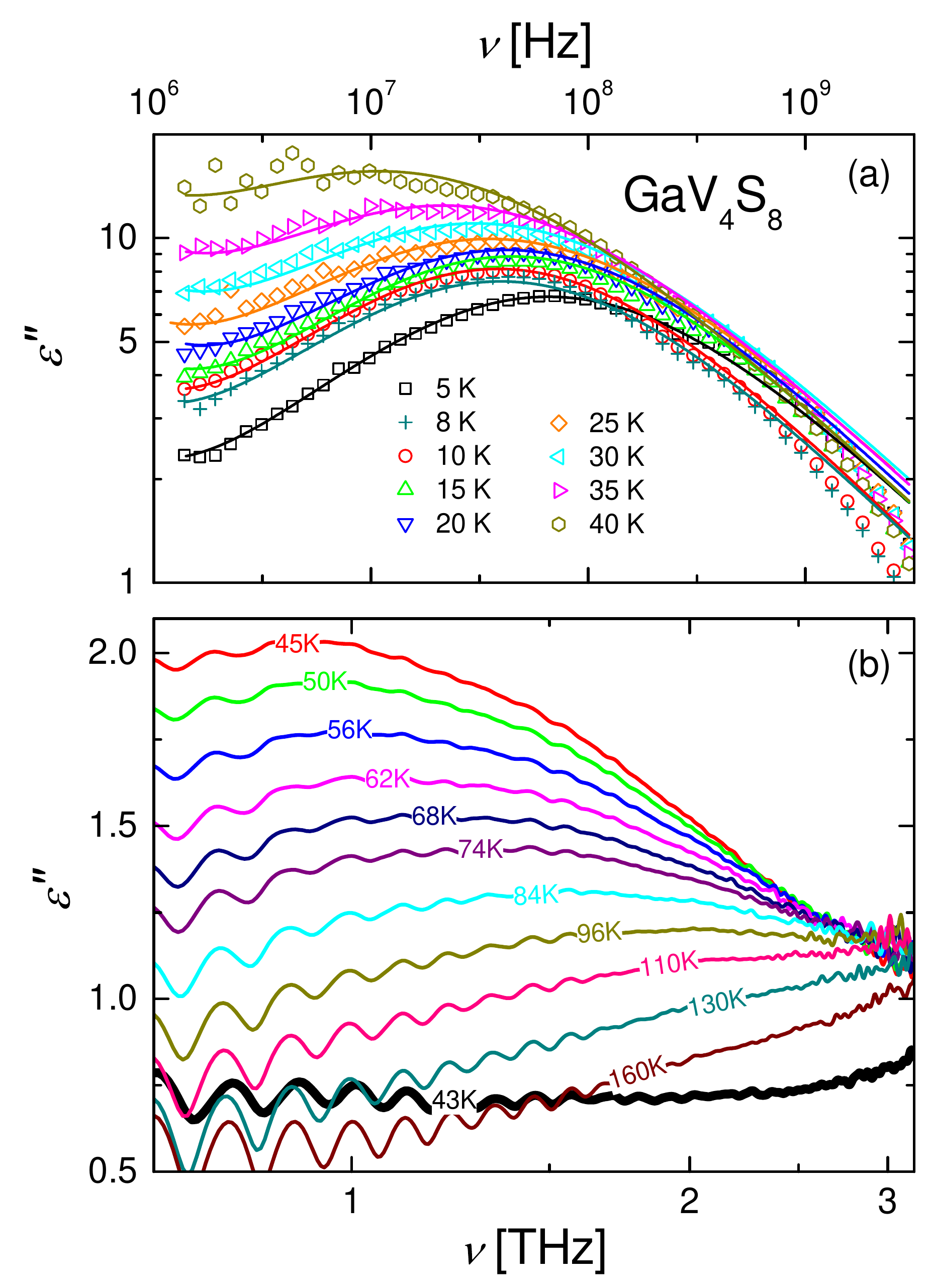}
  \caption{Dielectric loss spectra of \GVS\ \cite{wang2015}.  a) Spectra in the MHz - GHz frequency range for selected temperatures below \TJT. The lines show fits with the empirical Cole-Cole equation \cite{cole1941}. b) Spectra at 0.5 - 3~THz shown for temperatures from 160 to 43~K, the latter being just below the Jahn-Teller transition at 44~K (bold line). Adapted with permission.\textsuperscript{\cite{wang2015}} Copyright 2015, American Physical Society.}
  \label{Fig_Relax}
\end{figure}

Relaxation processes are best analyzed in frequency-dependent plots of the dielectric loss, $\varepsilon''(\nu)$, where they should lead to a peak when the angular frequency matches the inverse relaxation time \cite{lunkenheimer15,kremer2002}. As a typical example, the $\varepsilon''$ spectra of \GVS\ are shown in \textbf{Figure \ref{Fig_Relax}}a for various temperature below \TJT. The observed relaxation peaks shift to higher frequencies with decreasing temperature, opposite to the usual behavior of thermally activated processes \cite{lunkenheimer15,kremer2002}. Such an acceleration of the dynamics upon cooling just is in accord with the critical behavior expected below a order-disorder ferroelectric phase transition \cite{blinc1994,lines1996}. The loss peaks can be reasonably well fitted by the usual empirical functions used to describe dipolar relaxation processes \cite{kremer2002,cole1941} (lines in Figure \ref{Fig_Relax}a), providing information on the temperature-dependent mean relaxation time $\tau(T)$ discussed below. 

Interestingly, for \GVS, in the frequency range up to several GHz indications of intrinsic relaxation dynamics associated with dipolar motions were only found at $T<T_\mathrm{JT}$ \cite{ruff2015, wang2015, widmann2017}. However, clear relaxational response at $T>T_\mathrm{JT}$ was revealed in additional experiments performed at THz frequencies (Figure \ref{Fig_Relax}b) \cite{wang2015}. This implies that the corresponding relaxation time shifts by many decades when crossing the transition, which was ascribed to the first-order character of the orbital-ordering transition \cite{wang2015}. As shown in Figure \ref{Fig_Relax}b, with decreasing temperature the relaxation peak shifts into the experimental frequency window and moves to lower frequencies. Such canonical temperature dependence is indeed expected for order-disorder ferroelectrics at $T>T_\mathrm{JT}$ \cite{blinc1994,lines1996}. The peak amplitude continuously increases with decreasing temperature, which is related to the increase of $\varepsilon'(T)$ expected for ferroelectrics. As shown in ref. \cite{wang2015}, these spectra can be well fitted by the mono-dispersive Debye-relaxation function \cite{kremer2002}. 
While at 45\,K the peak is well pronounced, it abruptly vanishes below the transition as documented by the loss curve at 43\,K (bold line in Figure \ref{Fig_Relax}b), only 1~K below the phase transition. Here, the loss peak is shifted to lower frequencies, in accord with the low-frequency measurements, shown in Figure \ref{Fig_Relax}a. 

For \GVSe\ and \GNS, the dielectric spectra also reveal the typical signatures expected for dielectric relaxation processes \cite{geirhos2020b,ruff2017}. Interestingly, in antiferroelectric  \GNS\ both, below and above the Jahn-Teller transition relaxational behavior was detected in the low-frequency \cite{geirhos2020b} and the THz range as will be discussed below. For \GVSe, at $T>T_\mathrm{JT}$ non-intrinsic contributions prevent the detection of possible intrinsic relaxation processes in the low-frequency range while THz measurements have not yet been performed in this system.

\begin{figure}[h]
  \centering
  \includegraphics[width=0.6\linewidth]{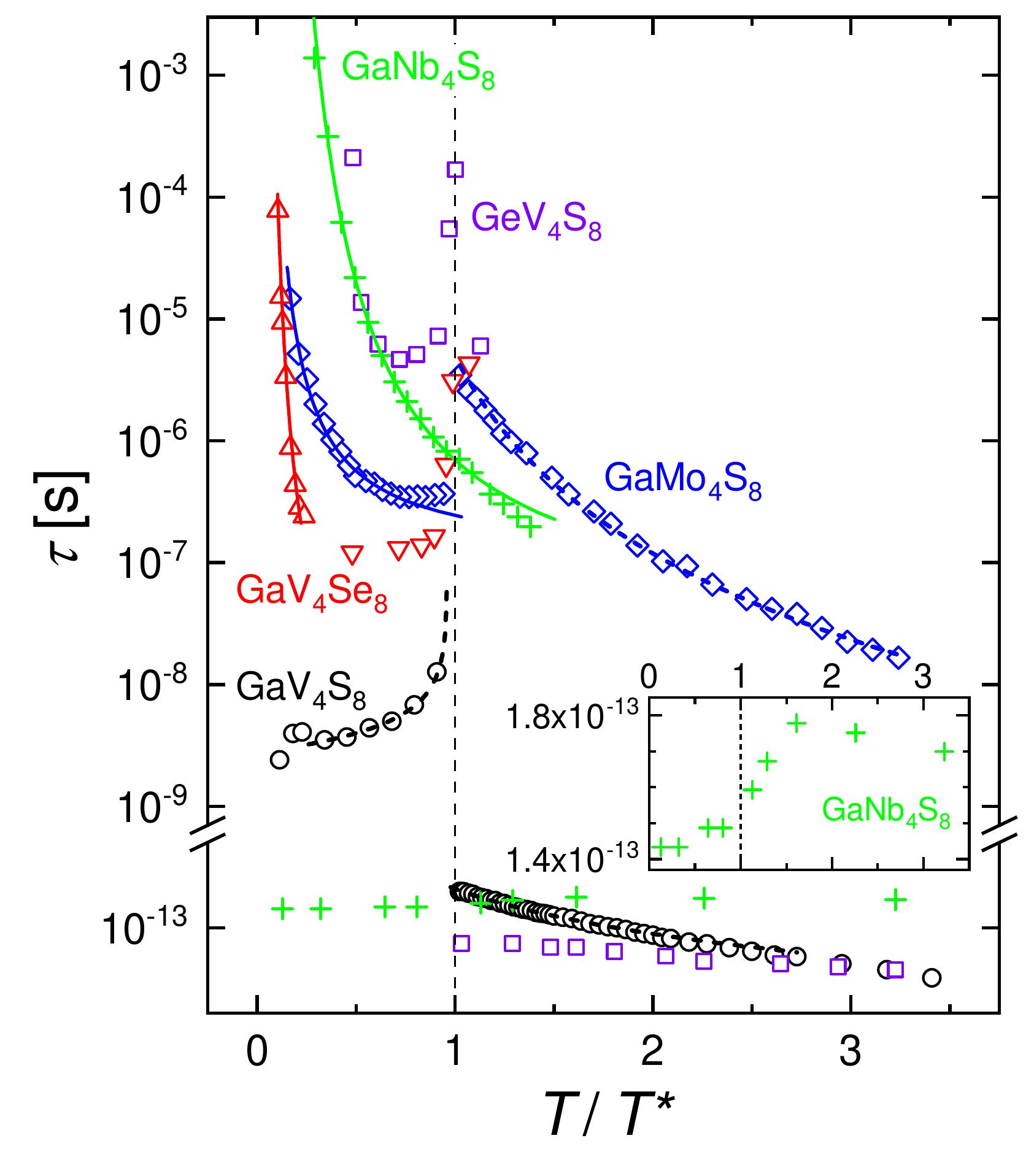}
  \caption{Overview of the temperature dependence of all relaxation times of dipolar relaxations found in lacunar spinels by dielectric and THz spectroscopy \cite{geirhos2018, geirhos2020b, wang2015, ruff2017, reschke2017b}. The temperature scale is normalized to the  $T^*$, which represents \TJT\ for each compound, except for \GNS  where it represents $T_\mathrm{MS}$ (see Table \ref{Table_Temperatures}). Solid and dashed lines represent fits of parts of the curves with Arrhenius or critical laws, respectively. The vertical dashed line indicates $T^*$. The inset shows an enlarged view of the \GNS\ fast-relaxation data.}
  \label{Fig_Taus}
\end{figure}

To quantify relaxational dynamics, usually relaxation times $\tau(T)$ are determined. They can be estimated from the frequency positions $\nu_p$ of the loss peaks via $\tau\approx1/(2\pi\nu_p)$ and determined by fits of the spectra \cite{lunkenheimer15,kremer2002}.
\textbf{Figure \ref{Fig_Taus}} shows the temperature dependence of $\tau$ of the three lacunar spinels discussed above and of \GMS\ and \GeVS, which will be treated in more detail below. Both above and below the Jahn-Teller transition, the relaxation times of these five systems partly differ by many decades and only some materials reveal common motifs as an increase of $\tau(T)$ when approaching the transition. These qualitative and quantitative differences could be expected in light of the different distortions of the cubane units found for most of these systems. However, even for \GVSe\ and \GVS\ with similar distortions, $\tau$ differs by more than a decade and its temperature dependence at the lowest temperatures is markedly different as discussed in more detail in the following paragraphs.

The triangles in Figure \ref{Fig_Taus} show the temperature dependence of $\tau$ of \GVSe. Here the upright triangles represent the mentioned region of conventional temperature characteristics of the relaxation dynamics, already indicated by the steps in $\varepsilon'(T)$ (Figure \ref{Fig_EpsT}a). Indeed, in this region at $T<9$~K, $\tau(T)$ can be well fitted by an Arrhenius law, with an energy barrier of 4.3~meV \cite{ruff2017} (solid line through the triangles). The inverted triangles in Figure \ref{Fig_Taus} show unpublished $\tau(T)$ data at higher temperatures, approaching \TJT. The increase of $\tau(T)$, observed when approaching the transition, is consistent with the expectation for order-disorder ferroelectrics \cite{blinc1994,lines1996}. The overall $\tau(T)$ curve at $T<T_\mathrm{JT}$, exhibiting a minimum, could thus be explained by critical behavior being relevant close to the transition only, while at lower temperatures thermal activation gains in importance and finally dominates the dynamics (see ref. \cite{ruff2017} for an alternative explanation). As mentioned above, at $T>T_\mathrm{JT}$, no information on the relaxation time of this system is available.

The temperature dependence of the relaxation time for \GVS\ is shown by the circles in Figure \ref{Fig_Taus}, making obvious the huge jump of $\tau(T)$ at \TJT\ by about five decades (please note the break of the $y$-axis) when the relaxation dynamics shifts into the THz range above \TJT\ as discussed above (cf. Figure \ref{Fig_Relax}). Similar to the typical second-order transitions of order-disorder ferroelectrics \cite{blinc1994,lines1996,hatta1968,gesi1970,deguchi1992,staresinic2006}, $\tau(T)$ increases when approaching the transition from both low and high temperatures and can be described by critical power laws, $1/(T-T_c)^\gamma$, in both cases (dashed lines) \cite{wang2015}. However, the ferroelectric transition in \GVS\ reveals several non-canonical features  as discussed in detail in ref. \cite{wang2015}. The huge jump in $\tau(T)$ probably can be ascribed to the clear first-order character of the transition \cite{widmann2017} and the fact that it is driven by the orbital order, however, the detailed microscopic origin  is far from being clarified. A close look at $\tau(T)$ of \GVS\ at the three lowest shown temperatures reveals some anomalies. They reflect the successive transitions into the cycloidal (deceleration of $\tau$) and the ferromagnetic state (acceleration of $\tau$). In this respect, \GVS\ behaves markedly different to \GVSe, discussed above, exhibiting strong thermally-activated relaxation dynamics in this low-temperature region.  However, notably the increase of $\tau(T)$ when approaching \TJT\ from below qualitatively resembles the behavior of \GVSe\ (inverted triangles). It seems interesting to measure the latter also in the THz range to check for a similarly strong acceleration of the relaxation dynamics when crossing \TJT.

The relaxation times of the antiferroelectric \GNS, deduced from the frequency-dependent dielectric data, are shown by the upper pluses in Figure \ref{Fig_Taus}. At $T<T_\mathrm{JT}$ they can be fitted by an Arrhenius law (solid line). Interestingly, in contrast to \GVSe\ and \GVS, $\tau(T)$ of \GNS\ does not exhibit any marked anomaly or critical behavior close to \TJT. Instead, it only crosses over to a different temperature dependence, deviating from the Arrhenius fit. In literature, information on dipolar fluctuations above and below the antiferroelectric phase transition is sparse. For antiferroelectric KCN, below the transition a relaxation process following Arrhenius behavior was observed, too \cite{luty83}. Similar to \GVS\ (cf. Figure \ref{Fig_Relax}b), additional investigations of \GNS\ in the THz region revealed a fast relaxation process which, however, also persists  below the Jahn-Teller transition, where it is superimposed by several sharp resonance excitations as will be reported in more detail in a future publication. Here, we only show its characteristic relaxation time, deduced from fits with the Debye formula (lower pluses in Figure \ref{Fig_Taus}), which appears in a similar range as for \GVS. However, $\tau(T)$ for \GNS\ is only weakly temperature dependent and exhibits a heavily smeared-out anomaly across the transition at best (inset of Figure \ref{Fig_Taus}). Thus, this process seems rather unrelated to the dipolar fluctuations associated with the antipolar transition and its physical origin still has to be clarified. 

Finally, we also briefly discuss the dielectric response of \GMS\ \cite{geirhos2018} and \GeVS\ \cite{widmann2016,reschke2017b}, two more lacunar spinels whose dielectric properties have been investigated in detail. As mentioned in section \ref{intro}, in \GVSe, \GVS, and \GNS, the dipole moments essentially arise from a stretching of the $M_4$ tetrahedra along one of their vertices, accompanied by a shrinking of the opposite face, which reduces the point-group symmetry of these units to $3m$. Although the change in point group of the $M_4$ units in \GMS\ is the same, there the vertex-face distance of the tetrahedra shrinks and the face area increases. This is due to the different electron occupations of the molecular-like orbitals in the $M_4$ clusters \cite{pocha2000}. This structural distortion also generates a dipole moment \cite{xu2015} and  \GMS\  becomes ferroelectric upon the Jahn-Teller transition to a rhombohedral phase  with switchable polarization below \TJT\ \cite{neuber2018, geirhos2018}. Thus, this lacunar spinel also is a type I multiferroic just as \GVSe\ and \GVS.

In a dielectric study of \GMS\ \cite{geirhos2018} complex relaxation dynamics was found. At $T<T_\mathrm{JT}$ a single intrinsic relaxation shows up, exhibiting thermally-activated temperature dependence at low temperatures and a possible transition to critical behavior close to \TJT, as shown by the diamonds and corresponding fit line in Figure \ref{Fig_Taus}. At $T>T_\mathrm{JT}$, two coupled intrinsic  dipolar-orbital relaxation processes were identified and ascribed to relaxor-ferroelectric-like dynamics of nanoregions with short-range polar order and to the critical fluctuations of weakly interacting dipoles, as expected for order-disorder ferroelectrics  \cite{geirhos2018}. The relaxation times of the latter, which can be fitted by a critical law (dashed line), are shown by the diamonds at $T>T_\mathrm{JT}$ in Figure \ref{Fig_Taus}. The relaxation behavior of \GMS\ above the Jahn-Teller transition markedly differs from that of \GVS\ where the dipolar dynamics at $T>T_\mathrm{JT}$ is much faster (circles in Figure \ref{Fig_Taus}). It seems likely that this is related to the mentioned differences in the structural distortions of the two lacunar spinels but the details are unclarified yet.  

\GeVS\ was the first lacunar spinel whose dielectric properties were investigated in detail, revealing orbital-order driven multiferroicity \cite{singh2014}. In contrast to all the lacunar spinels treated above, it has two unpaired electrons with total spin $S=1$ in the highest orbital level of the $M_4$ tetrahedra. This leads to a significantly different distortion of these units via an elongation and shortening of two opposite V-V bonds, respectively, and an overall symmetry change to the space group $Imm2$ upon \TJT~\cite{johrendt1998, muller2006,bichler2008}.  As first detected by Singh \textit{et al.} \cite{singh2014}, the behavior of the dielectric constant of \GeVS\ differs markedly from that of the  compounds discussed above \cite{widmann2016,reschke2017b}: Below \TJT, $\varepsilon'(T)$ strongly increases and a broad maximum develops several Kelvin below the transition. At the antiferromagnetic ordering a further structural transition takes place and $\varepsilon'(T)$ is strongly suppressed \cite{singh2014}, while the other ferroelectric systems do not exhibit significant anomalies at their magnetic transitions. 
An additional peak close to \TJT\ was detected in ref. \cite{widmann2016} and suggested to indicate a possible decoupling of orbital and polar phase transition. A detailed dielectric study at audio and radio frequencies evidenced a relaxation process whose amplitude is strongly reduced above \TJT\ and in the antiferromagnetic phase \cite{reschke2017b}. It was proposed to be due to the coupled orbital and dipolar fluctuations that are associated with the polar order \cite{reschke2017b}. The corresponding relaxation times are shown by the upper squares in Figure \ref{Fig_Taus} revealing a slowing down when approaching both the structural \emph{and} the magnetic transition, the latter in contrast to the other lacunar spinels. Additional dielectric THz measurements were also performed for this compound, evidencing a number of different excitations \cite{reschke2017b}. Among them is a Debye-like relaxation process that appears at $T>T_\mathrm{JT}$. Its relaxation times are shown by the lower squares in Figure \ref{Fig_Taus}. This fast process reminds of the THz relaxation reported for \GVS\ (lower circles in Figure \ref{Fig_Taus}), reflecting dipolar fluctuations related to the order-disorder ferroelectric transition \cite{wang2015}. However, in contrast to \GVS, below the Jahn-Teller transition the THz relaxation of \GeVS\ was reported to seemingly remain located in the THz regime while its character crosses over from relaxational to resonance-like \cite{reschke2017b}. Therefore, there may be no close relation of this fast relaxation dynamics to the one detected below \TJT\ in the dielectric frequency range.

Overall, the above overview about the dielectric properties of five lacunar spinels, four of them being multiferroic, demonstrates that there is a large variation in the rather complex dielectric response of these materials showing non-canonical, orbital-driven polar order. In fact, the dipolar moments in these compounds are in principle generated by the same mechanism, namely a polar distortion of the molecule-like $M_4$ units. However, partly due to variations in the occupation of their electronic energy levels, these distortions qualitatively differ for most of these systems. Common to all investigated lacunar spinels is the occurrence of relaxational dynamics. This points to the order-disorder character of their polar transitions involving coupled dipolar and orbital fluctuations, consistent with a dynamic Jahn-Teller effect. A common motif of the relaxation-time curves shown in Figure \ref{Fig_Taus} (except for the antipolar \GNS) is the increase of the relaxation time when approaching the polar transition, which partly could be fitted by critical laws. This finding is in accord with the order-disorder scenario. At the lowest temperatures, thermal activation starts to dominate the coupled dipolar-orbital motions in all systems except \GVS, leading to a more or less pronounced minimum in $\tau(T)$. Many open questions remain to be clarified concerning the partly huge differences in the relaxation times of these materials and, especially, the unusual many-decades jump of $\tau(T)$ that seems to arise at the transition of some of these non-canonical polar systems.

\section{Ferro- and magnetoelectric polarization} \label{MulMag}

As shown in the previous chapter, a pronounced dielectric anomaly at \TJT\ hints to the onset of long-range polar order in various lacunar spinels. The emergence of ferroelectricity was directly evidenced by temperature- and magnetic field-dependent polarization data, respectively obtained via pyro- and magnetocurrent measurements, in \GeVS~\cite{singh2014}, \GVS~\cite{ruff2015}, \GVSe~\cite{fujima2017, ruff2017}, and \GMS~\cite{neuber2018}. Due to their spontaneous polarization and type-I multiferroic nature, these materials offer an ideal playground to investigate i) novel magnetoelectric effects \cite{ruff2015, singh2014,  fujima2017, geirhos2020,felea2020,  nikolaev2019, nikolaev2020}  and ii) ferroelectric domain and domain-wall properties \cite{neuber2018, geirhos2020, kezsmarki2015, butykai2017}.   

Singh \textit{et al.} \cite{singh2014} provided the first comprehensive study evidencing  orbital-order driven ferroelectric behavior and  magnetoelectric coupling of a lacunar spinel, namely \GeVS. Recently, the ferro- and magnetoelectric properties of the rhombohedral lacunar spinels \GVS , \GVSe, and \GMS\  have also been studied in detail. These compounds exhibit unique polar properties, such as the lack of inversion domains, full magnetic control of polar states (see section~\ref{pol}), and N\'eel-type magnetic skyrmions dressed with polarization (see section \ref{MagPol}) \cite{ruff2015,  fujima2017,  xu2015,neuber2018, geirhos2020, ruff2017, nikolaev2019, nikolaev2020, zhang2019}.

\begin{figure}[h]
\centering
  \includegraphics[width=0.5\linewidth]{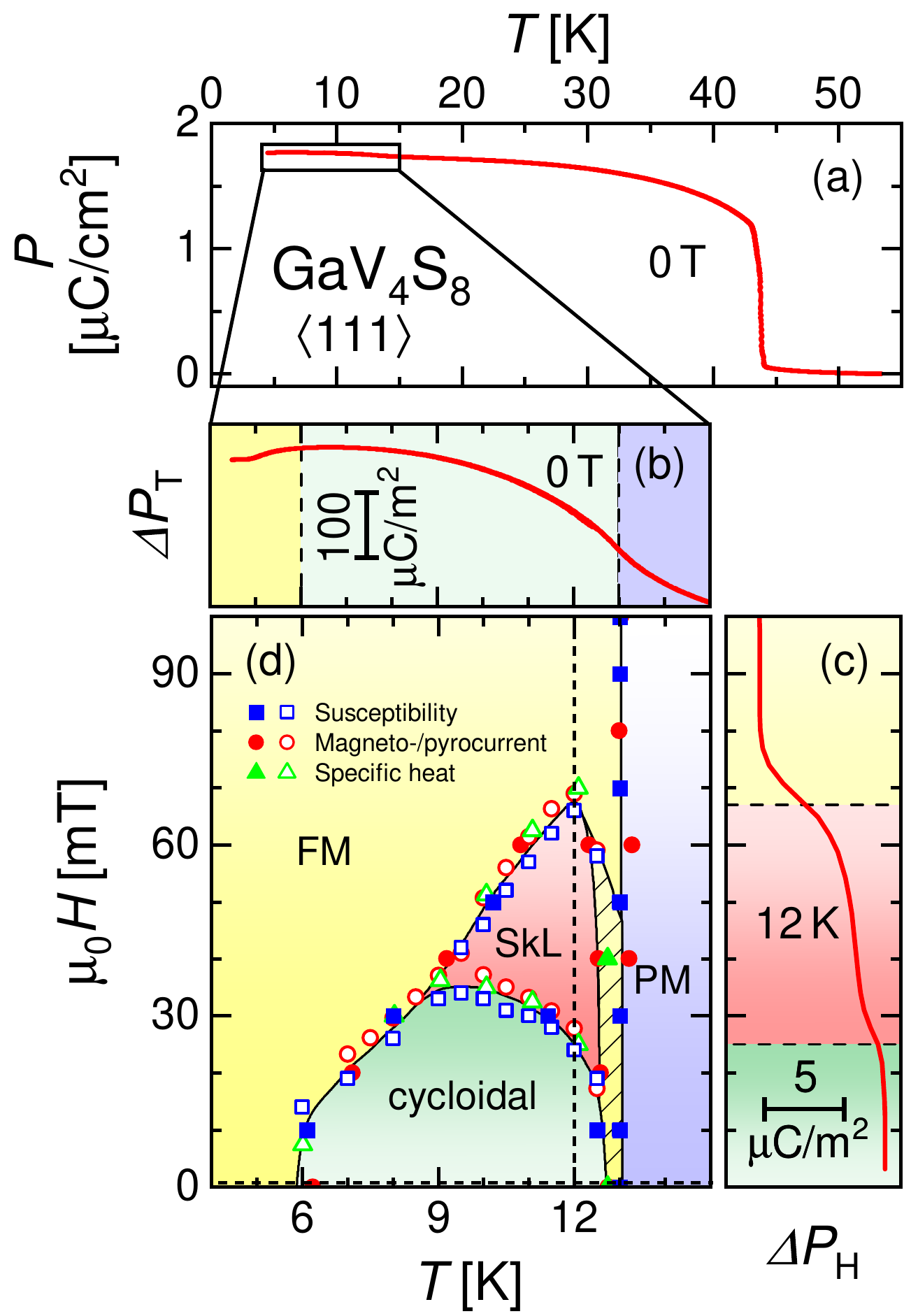}
  \caption{a) Temperature-dependent polarization of \GVS .  b) Temperature-dependent spin-induced polarization at 0\,T observable in the temperature range marked by the box in a. c) Magnetic-field-dependent spin-induced polarization at $12\,$K \cite{ruff2015}. d) Magnetic phase diagram of \GVS\ \cite{ruff2015}. The horizontal and vertical dashed lines in d indicate the locations of b and c, respectively. Adapted under terms of the CC-BY license.\textsuperscript{\cite{ruff2015}} Copyright 2015, The Authors, published by American Association for the Advancement of Science.}
  \label{Fig_Polarization}
\end{figure}

Ruff \textit{et al.} \cite{ruff2015} provided a thorough characterization by magnetic susceptibility, specific heat, as well as magneto- and pyrocurrent measurements of the polar properties of \GVS. \textbf{Figure~\ref{Fig_Polarization}}a shows typical temperature-dependent polarization data measured along the $\langle111\rangle$-type axis. 
It is characterized by a sudden jump from zero to finite values below $T_\mathrm{JT}=44$\,K followed by a moderate increase, which is typical for the order parameter of a first-order transition. The weak first-order nature of the ferroelectric transition was also corroborated by specific-heat measurements \cite{widmann2017}. Similar polarization results were also reported for \GVSe~\cite{fujima2017, geirhos2020, ruff2017} and \GMS~\cite{neuber2018}. 

As illustrated by Figure \ref{Fig_Polarization}a and b, a second increase in the polarization shows up at $T_\mathrm{C}$ = 13\,K, as a result of an excess spin-induced polarization upon the magnetic ordering, indicating the presence of magnetoelectric coupling \cite{ruff2015, fujima2017, geirhos2020, ruff2017}. The spin-induced polarization rises when entering the cycloidal spin state at $T_\mathrm{C}$, and is somewhat reduced when entering the ferromagnetic state at 6\,K. Interestingly, the magnetoelectric polarization distinguishes all the magnetically ordered states, i.e., the excess polarization has three distinct values in cycloidal, N\'eel-type skyrmion lattice, and ferromagnetic states (see magnetic phase diagram in Figure~\ref{Fig_Polarization}d), as depicted in Figure~\ref{Fig_Polarization}c. This pioneering finding evidences that N\'eel-type skyrmions can also carry ferroelectric polarization \cite{ruff2015}, which may allow the electric control and readout of N\'eel-type skyrmions. 

Spin-induced polarization was also observed in \GVSe~\cite{fujima2017, geirhos2020, ruff2017}, which we discuss in detail in section~\ref{MagPol}, on the basis of original data. In \GVSe, the skyrmions also carry electric polarization and the skyrmion lattice region in the phase diagram is strongly extended as compared to \GVS, namely it is stable down to zero Kelvin. The thermal and field stability ranges of the skyrmion lattice in \GVS\ and \GVSe\ have been studied in great details experimentally \cite{geirhos2020, kezsmarki2015, bordacs2017, gross2020} and theoretically \cite{zhang2019, leonov2017, leonov2017b, kitchaev2020}.  For the rhombohedral lacunar spinel \GMS, such a spin-induced polarization has been predicted by Nikolaev \textit{et al.} \cite{nikolaev2020} but not yet observed, making this system also highly interesting. 

From a more applied point of view, one important aspect of ferroelectrics is the control of the polarization by external stimuli, such as electric and/or magnetic fields. Despite the observation of polarization-dressed skyrmions and remarkable magnetoelectric couplings in these orbital-ordered ferroelectric systems, the on-demand control of the ferroelectric state remains a nearly uncharted territory in ferroelectric lacunar spinels.
The control of the ferroelectric polarization by electric fields was first demonstrated and discussed in detail only for \GeVS\ \cite{singh2014}. In case of the ferroelectric lacunar spinels with rhombohedral structure, the electric-field manipulation of polar domains was reported in \GVSe\  \cite{geirhos2020} and \GMS\  \cite{neuber2018}. So far, ferroelectric polarization control has not been demonstrated in case of the first reported N\'eel-type skyrmion host, \GVS. Here, we choose \GVSe\ to emphasize the feasibility of an electric-field control of magnetic skyrmions and to elucidate prospects for future investigations. In the following section, we describe the polarization switching by external magnetic and electric fields in this compound.

\subsection{Electric and magnetic control of domain states} \label{pol}

\begin{figure}[h]
\centering
  \includegraphics[width=0.9\linewidth]{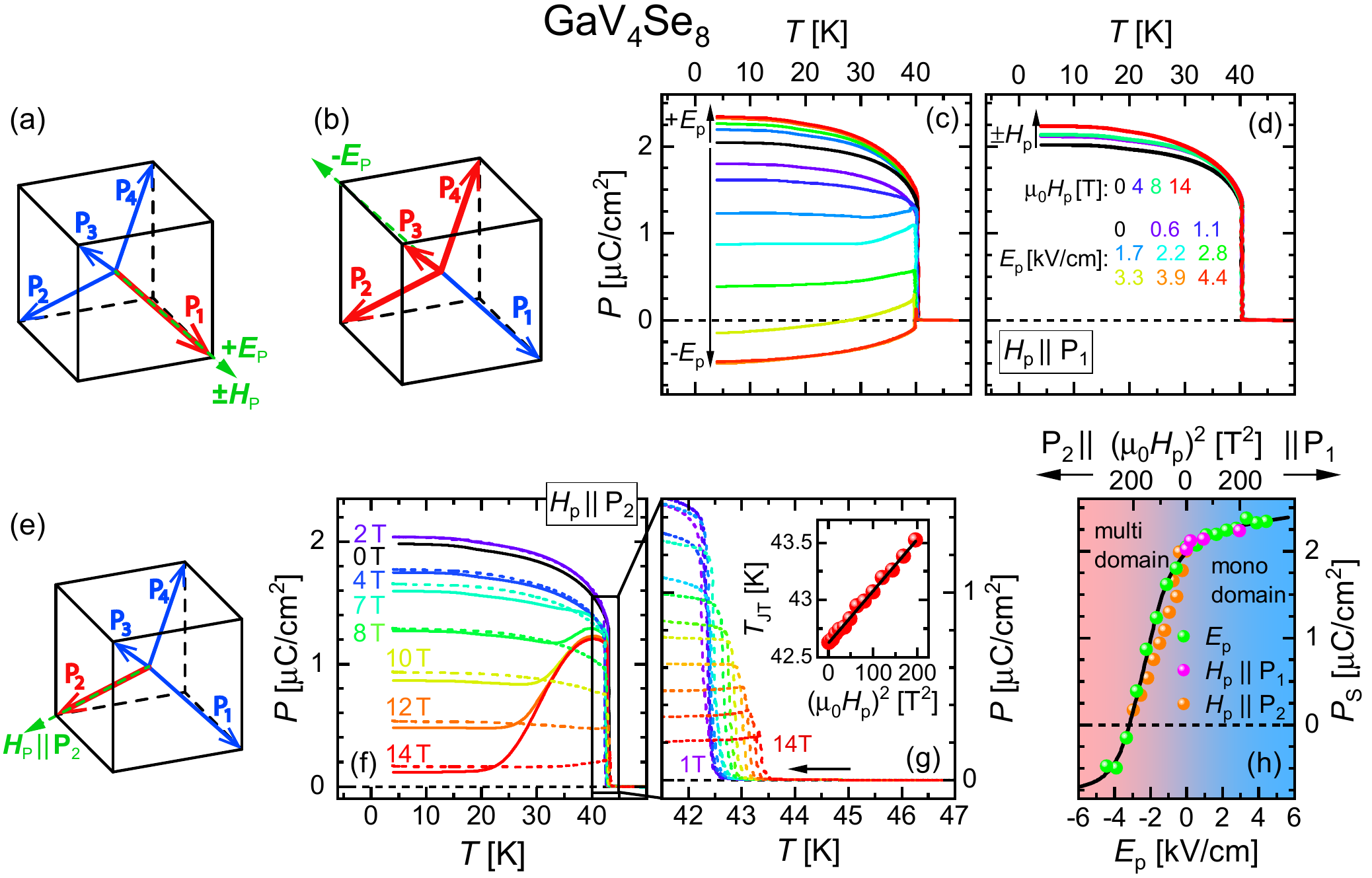}
  \caption{
a) \& b) Schematic representations of the domain-selection process by electric and magnetic fields applied along the $P_1$ direction in the rhombohedral lacunar spinels. c) \& d) Temperature-dependent polarization of \GVSe\ for various electric and magnetic poling fields applied along [111] ($P_1$), respectively. e) Schematic representation of the direction of applied field used for the measurements shown in f and g. f) Temperature-dependent polarization for various strengths of magnetic poling fields applied along $P_2$ measured during cooling (dashed lines) and heating (solid lines). g) Magnified view of the cooling curves in the area marked in f (additional fields included). The inset in g displays the magnetic field dependence of \TJT . h) Electric and magnetic poling field dependence of the saturation polarization $P_\mathrm{s}$ at 4\,K. In all shown measurements the polarization $P$ was measured along $P_1$.}
  \label{Fig_Poling}
\end{figure}

As discussed in the introduction, the room-temperature cubic structure of lacunar spinels already lacks inversion symmetry \cite{johrendt1998, barz1973}. As a consequence, in compounds undergoing a transition from the cubic state to the rhombohedral state ($R3m$), four polar domain states ($P_1$, ..., $P_4$) can form with their polar axes pointing along the four cubic body diagonals ($[111]$, $[1\bar{1}\bar{1}]$, $[\bar{1}1\bar{1}]$, $[\bar{1}\bar{1}1]$, as shown in Figure~\ref{Fig_TransTemp}b \cite{geirhos2020, kezsmarki2015, butykai2017}). A unique feature of this ferroelectric state is the lack of inversion domains ($-P_1$, ..., $-P_4$). This polar-diode nature enables a peculiar control of the domain states by external fields. The application of an electric field ($+E_\mathrm{p}$) along a polar axis, e.g. $P_1$, promotes a single domain state, namely the $P_1$ domain state (see \textbf{Figure~\ref{Fig_Poling}}a), while an electric field of opposite sign ($-E_\mathrm{p}$) equally favors the other three domains $P_2$, $P_3$ and $P_4$ (see Figure~\ref{Fig_Poling}b), thus, enforcing the creation of a polar multi-domain state.

Figure~\ref{Fig_Poling}c shows the manipulation of the ferroelectric polarization for \GVSe\ by the application of various electric poling field ($E_\mathrm{p}$) applied along [111] axis, demonstrating the electric control of the domain population. The polarization component parallel to the [111] axis is derived from temperature-dependent pyrocurrent measurements (for experimental details see ref.~\cite{geirhos2020}). The temperature-dependent polarization resembles that of \GVS\ (see Figure~\ref{Fig_Polarization}a). 

It is interesting that a significant polarization of the order of $2\,\mu\mathrm{C}/\mathrm{cm}^2$ is observed below \TJT\, even without any poling field. This hints towards a natural preference on the domain state, $P_1$ in the present case, with its polar axis perpendicular to the planar metallic electrodes. Increasing the positive poling field $E_\mathrm{p}$ applied along the direction of $P_1$ up to 4.4\,kV/cm leads to a slight enhancement and a saturation of the polarization to $P_\mathrm{mono}\sim 2.3\,\mu\mathrm{C}/\mathrm{cm}^2$. This saturation indicates the achievement of a polar mono-domain state. Applying poling fields of opposite sign ($-E_\mathrm{p}$) results in a gradual reduction of the macroscopic polarization with increasing strength of $-E_\mathrm{p}$. This is due to the decreasing population of the $P_1$ domain state and the enhanced population of the other three domains, as sketched in Figure \ref{Fig_Polarization}b. For $E_\mathrm{p}$=$-4.4\,\mathrm{kV}/\mathrm{cm}$ the polarization measured along the [111] axis turns to a negative value, indicating that the volume fraction of the $P_1$ domain state is less than 50\%. When the $P_1$ domain state is fully suppressed by negative poling electric fields the polarization is expected to saturate to $P_\mathrm{mono}/3\sim-0.7\,\mu\mathrm{C}/\mathrm{cm}^2$. 

A direct comparison of the absolute value of polarization among different lacunar spinels is challenging, as either the relative populations of the four domains has to be known or a mono-domain state has to be achieved. The polarization has only been measured on multi-domain samples of \GVS\ \cite{ruff2015} and only estimated for \GMS\ \cite{neuber2018}. The polarization in the mono-domain state is only known for \GVSe\, where the saturation value is $P_\mathrm{mono}\sim 2.3\,\mu\mathrm{C}/\mathrm{cm}^2$, as seen in Figure \ref{Fig_Polarization}c and d. From theoretical analysis, polarization values calculated using a Berry phase approach are reported to be $2.43\,\mu\mathrm{C}/\mathrm{cm}^2$ for \GVS\ \cite{xu2015} and $\sim3.1\,\mu\mathrm{C}/\mathrm{cm}^2$ for \GMS~\cite{zhang2019}. Notably, the experimental value for \GVSe\ agrees well with the one calculated for \GVS . 

Interestingly, the ferroelectric polarization of \GVSe\ can be controlled not only by electric but also by magnetic fields ($H_\mathrm{p}$). The realization of a mono-domain state is possible via cooling the material through \TJT\ in magnetic fields applied along one of the cubic body diagonals, in the present case the $P_1$ direction (see Figure~\ref{Fig_Poling}d). The magnetic control of the ferroelectric domain population is facilitated by the uniaxial magnetic anisotropy of the material. Since the anisotropy axis coincides with the polar axis, the domain with easy-axis along the field is favoured \cite{kezsmarki2015, bordacs2017, ehlers2016b}. Due to the fact that the magnetic anisotropy energy is an even function of the magnetic field, the same domain state is favored for both positive and negative signs of the field.

The question arises, whether any of the four mono-domain states can be established, although the samples usually show a preference on one domain state due to the contact geometry or internal strain. For this, we apply poling magnetic fields along a polar axis, different from the naturally favoured one, namely along $P_2\parallel$[$1\bar{1}\bar{1}$], as sketched in Figure~\ref{Fig_Poling}e. Figure~\ref{Fig_Poling}f shows the temperature-dependent polarization recorded along $P_1$, the polarization direction of the originally favoured domain, for various poling magnetic fields applied along $P_2$. As clear from Figure~\ref{Fig_Poling}f, with increasing magnetic field the polarization reduces continuously to values of the order of $\sim 0.1\,\mu\mathrm{C}/\mathrm{cm}^2$ at 4\,K with a maximum applied field of $\mu_0H_\mathrm{p}=$ 14\,T. The nearly zero net polarization in 14\,T indicates that the volume fraction of the magnetically promoted $P_2$ domain state is about 3 times larger than the remaining fraction of the $P_1$ domain state. Comparing the temperature dependence of the polarization from pyrocurrent measurements after poling with a magnetic field along $P_1$ (Figure~\ref{Fig_Poling}d) and along $P_2$ (Figure~\ref{Fig_Poling}f, solid lines) reveals another interesting aspect: For poling magnetic fields $\mu_0H_\mathrm{p}> 7\,$T along $P_2$, a strong enhancement of the polarization is observed above 25\,K, before vanishing at \TJT . Since the magnetic field is only used for the poling and switched off during the polarization measurement, performed in warming up, this feature indicates a temperature induced back-switching from the state dominated by $P_2$ domains to the naturally preferred $P_1$ domain state. As already shown, via negative poling electric fields applied along the [111] axis, one can also achieve a multi-domain state where the fraction of the $P_1$ domain state is suppressed. However, in that case, a temperature induced back-switching to $P_1$ domain state, emerging upon zero-field heating, does not take place, as clear from Figure~\ref{Fig_Poling}c.  

To verify the back-switching mechanism, we recorded the temperature-dependent polarization during cooling in applied magnetic fields of up to 14\,T. The results are indicated by dashed lines in Figure~\ref{Fig_Poling}f. These curves resemble the behavior seen in the electric-field poling measurements (Figure~\ref{Fig_Poling}c).

In addition to the domain control, external fields can also shift \TJT\ towards higher temperatures, as illustrated in Figure~\ref{Fig_Poling}g, which is a magnified view of the box indicated in Figure \ref{Fig_Poling}f. At 14\,T, the ferroelectric transition temperature is increased by $\sim1$\,K. This shift is most likely caused by the higher magnetic susceptibility of the rhombohedral state compared to the cubic state \cite{ruff2017}. This explanation is also supported by the quadratic magnetic-field dependence of the transition temperature, which is obvious from the inset in Figure~\ref{Fig_Poling}g. Due to technical limitations, recording the ferroelectric polarization upon cooling in the presence of electric fields is not possible in this compound, thus, the electric field-induced shift of \TJT\ could not be investigated.

Finally, Figure~\ref{Fig_Poling}h summarizes the thorough electric and magnetic field control of the domain states in \GVSe . The net saturation polarization is plotted as function of poling electric and magnetic fields. The latter also takes into account the dependence of domain population on the orientation of the applied magnetic field ($H_\mathrm{p}\parallel P_1$ and $P_2$). The polarization measured at the lowest temperatures $P_\mathrm{s}$ depends on the poling electric field as depicted by the green circles. A nearly complete control from a mono- to a multi-domain state can be achieved. Also magnetic fields enable the control of the domain states. In this case, however, the saturation polarization has a quadratic magnetic-field dependence, as expected from the field dependence of the magnetic free energy. Resulting from this quadratic dependence, one domain state is favored when applying magnetic fields along any of the polar axes, irrespective of the sign of the field. Moreover, the selection of different mono-domain states can be achieved by changing the orientation of the poling magnetic field.

\subsection{Spin-driven polarization} \label{MagPol}

For \GVS\ it was shown by Ruff \textit{et al.} \cite{ruff2015} that all magnetic phases emerging at low temperatures, namely a cycloidal, a N\'eel-type skyrmion lattice, and a field-polarized ferromagnetic phase, carry different spin-induced ferroelectric polarizations, as illustrated in Figure~\ref{Fig_Polarization}. The other rhombohedral lacunar spinels, \GVSe\ and \GMS, also undergo magnetic transitions at $T_\mathrm{C}$ = 18\,K \cite{bordacs2017} and $T_\mathrm{C}$ = 20\,K \cite{butykai2019}, respectively. They were also found to exhibit complex magnetic phase diagrams. Bordacs \textit{et al.} \cite{bordacs2017} detected a N\'eel-type skyrmion lattice state in \GVSe , which was corroborated by Fujima \textit{et al.} \cite{fujima2017} to exhibit a spin-induced polarization similar to \GVS~\cite{ruff2017}. In case of \GMS , the existence of a skyrmion lattice \cite{butykai2019, zhang2019, kitchaev2020} as well as magnetoelectric coupling \cite{nikolaev2020} have been proposed, but not experimentally verified yet.

The uniaxial magnetic anisotropy, inherent to the rhombohedral state, leads to the coexistence of distinct magnetic phases on different polar domains for magnetic fields of arbitrary orientation, which makes the determination of magnetic phase diagrams a delicate issue. These difficulties can be overcome by studying the magnetic phase diagrams of these compounds on rhombohedral mono-domain crystals, achieved by electric and/or magnetic poling, as demonstrated for \GVSe\  \cite{geirhos2020}. Another approach, which has been applied to rhombohedral multi-domain samples of \GVS\ \cite{kezsmarki2015} and \GVSe\ \cite{geirhos2020,gross2020}, is tracing the angular dependence of the critical magnetic fields. As a very important outcome, these studies reveal highly anisotropic magnetic phase diagrams of these materials. While the cycloidal state is always stable in low fields, irrespective of the orientation of the magnetic field, the stability of the skyrmion lattice state is limited to a critical angle of oblique fields, i.e., it is stable for fields applied along the rhombohedral axis but vanishes if the magnetic field is tilted beyond a critical angle \cite{geirhos2020, kezsmarki2015, gross2020}. The highly anisotropic nature of the magnetic phase diagrams has been reproduced theoretically in these compounds \cite{zhang2019, leonov2017, kitchaev2020}.

\begin{figure}
\centering
  \includegraphics[width=0.5\linewidth]{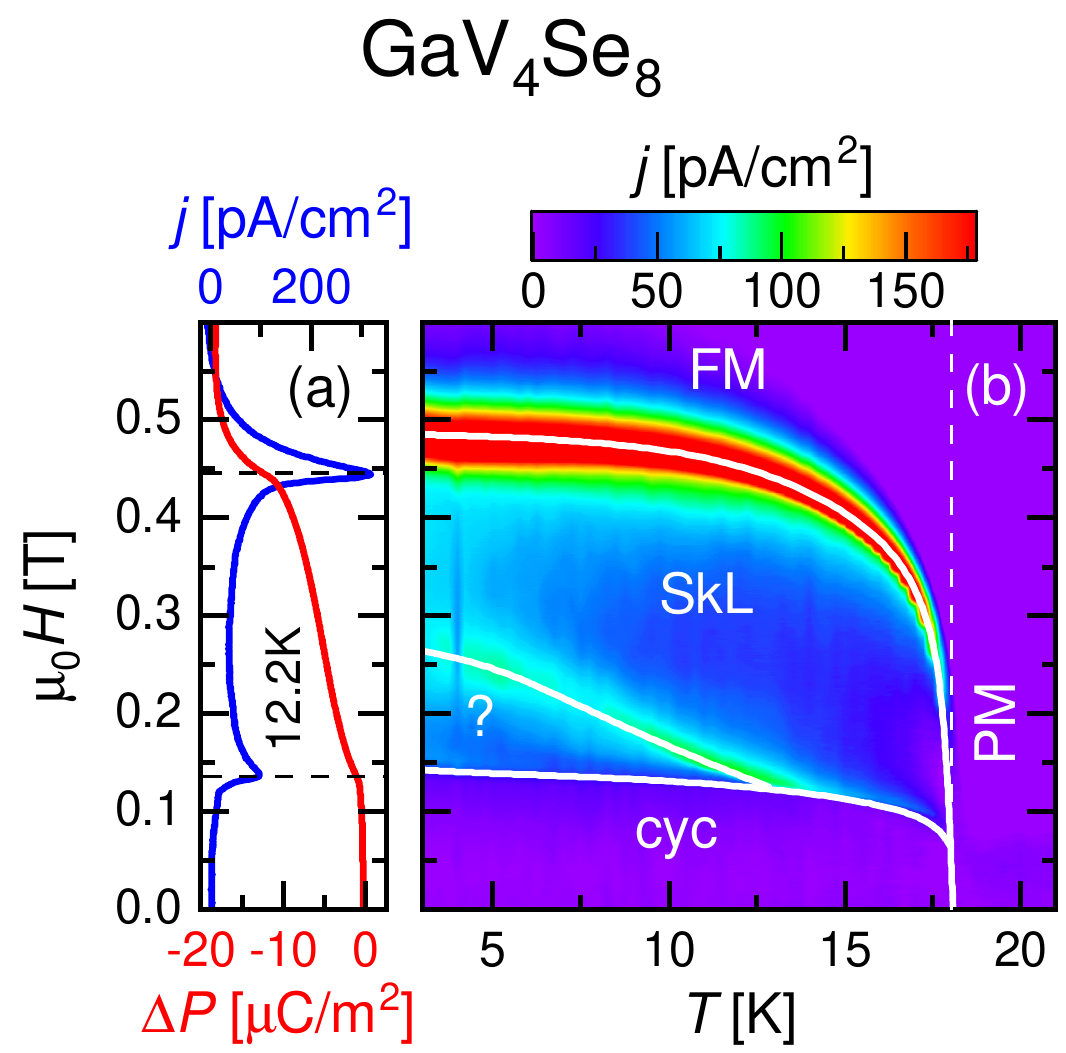}
  \caption{ a) Magnetic-field-dependent magnetocurrent and  polarization at $12.2$\,K of a mono-domain \GVSe\ crystal along $\langle111\rangle$ with magnetic field oriented along the same direction. b) Color map of the full temperature and magnetic-field dependence of the magnetocurrent measured with the same configuration as in a.}
  \label{Fig_MagnetoPol}
\end{figure}

Here, we briefly describe through the example of \GVSe\ how the progress in domain control helps exploring the magnetic phase diagram via the detection of spin-induced polarization, which exhibits clear anomalies at the magnetic transitions similarly to the magnetic susceptibility. This was first demonstrated in \GVS\, where peaks in the magnetocurrent and corresponding steps in the polarization have been assigned to transitions between the different magnetic phases \cite{ruff2015}, allowing the determination of stability ranges of the cycloidal, N\'eel-type skyrmion lattice and ferromagnetic states. \textbf{Figure~\ref{Fig_MagnetoPol}}a shows the magnetocurrent ($j$, time derivative of polarization) and the corresponding polarization on a mono-domain \GVSe\ crystal at $T=12.2$\,K. Anomalies observed in the color map of the magnetocurrent over the temperature--field plane clearly reveal the magnetic phase boundaries, as seen in Figure~\ref{Fig_MagnetoPol}b. In this compound, we observed four magnetic phases, all of them exhibiting a distinct spin-induced polarization. Small-angle neutron scattering measurements \cite{geirhos2020, bordacs2017} allow to ascribe three phases to cycloidal, N\'eel-type skyrmion lattice, and ferromagnetic phase, as indicated in Figure~\ref{Fig_MagnetoPol}b. The precise spin texture of the fourth phase, emerging only below 12\,K between the cycloidal and skyrmion lattice phases, is still unknown. Interestingly, the skyrmion lattice phase of \GVSe\ is strongly extended in the temperature range compared to \GVS\ (see Figure \ref{Fig_Polarization}d), which is explained by a weaker magnetic anisotropy of \GVSe~\cite{fujima2017, bordacs2017}. This study, carried out on a mono-domain crystal, precisely quantifies the magnetoelectric effect in all magnetic phases of \GVSe .   

Though such a detailed systematic study of the magnetoelectric effect has not been performed in the other rhombohedral lacunar spinels, the spin-induced polarization varies in the same order of magnitude for all compounds \cite{ruff2015, fujima2017, geirhos2020}. From a technical point of view, this spin-induced polarization arising in rhombohedral lacunar spinels is about two orders of magnitude larger compared to that of Cu$_2$OSeO$_3$, another magnetoelectric insulator hosting Bloch-type skyrmions \cite{seki2012, ruff2015b}. In that system, a quadratic magnetoelectric effect is discussed as the origin of the polarization of modulated magnetic phases, including the Bloch-type skyrmion phase \cite{seki2012, ruff2015b}. In contrast, the excess polarization in the skyrmion lattice phase of \GVS\ was proposed by Ruff \textit{et al.} to originate from the exchange-striction mechanism \cite{ruff2015}, which leads to a polar ring around the core of the N\'eel-type skyrmions. The finding of N\'eel-type skyrmions carrying an inherent electric polarization could pave the way towards the development of novel energy-efficient electric-field controlled skyrmion-based electronics.

\section{Outlook: Towards multiferroic domain and domain wall properties}

As overviewed in the present work, the optical, dielectric, and polar properties of numerous lacunar spinel compounds have been thoroughly investigated and discussed in the literature. Optical studies classify all the investigated lacunar spinels as narrow gap semiconductors. As another common aspect, all the investigated compounds exhibit a structural transition, which lowers the symmetry of the lattice from cubic to rhombohedral or orthorhombic. Depending on the compound, this transition can lead to either a ferroelectric or  an antiferroelectric low-temperature state. Due to the presence of dielectric relaxations, evidencing coupled dipolar and orbital fluctuations, the polar transition in all compounds can be classified as order-disorder type. This implies the presence of a dynamical Jahn-Teller effect in the cubic phase. Still, many open questions remain to be clarified concerning the non-canonical dielectric behavior of these unconventional ferro- and antiferroelectric materials. A further general aspect of these compounds is the spin-lattice coupling: In case of ferroelectric lacunar spinels this is manifested by a change of the exchange interaction from antiferromagnetic (cubic phase) to ferromagnetic (rhombohedral phase), while in the antiferroelectric compounds the symmetry-lowering structural transition is associated with a spin-singlet formation. The ferroelectric compounds exhibit a sizable polarization, which can be controlled by electric, as well as magnetic fields. Moreover, the magnetic order, taking place within the polar state renders these compounds type-I multiferroics, which show a decent magnetoelectric coupling.

\begin{figure}
\centering
  \includegraphics[width=0.9\linewidth]{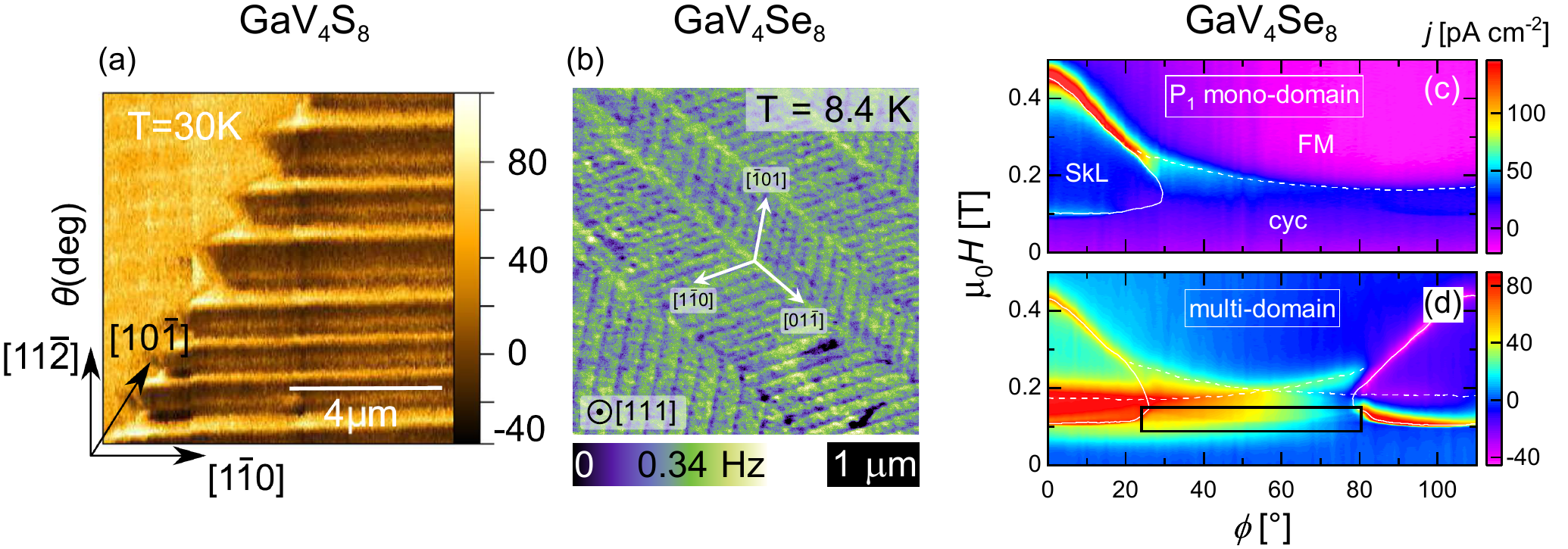}
  \caption{a) Phase of a PFM image recorded on a (111) surface of \GVS~ \cite{butykai2017}. Reproduced under terms of the CC-BY license.\textsuperscript{\cite{butykai2017}} Copyright 2017, The Authors, published by Springer Nature. b) Frequency shift of a non-contact MFM image recorded on \GVSe~ \cite{geirhos2020}. Note that the contrast in this image is not a magnetic one, but stems from the polar domain structure \cite{geirhos2020}.  Reproduced under terms of the CC-BY license.\textsuperscript{\cite{geirhos2020}} Copyright 2020, The Authors, published by Springer Nature. c) \& d) Color maps of the magnetoelectric current-density as a function of magnetic-field angle and magnitude measured on \GVSe\ at 12\,K in the polar mono- and multi-domain state, respectively \cite{geirhos2020}. The lines mark anomalies belonging to magnetic transitions in the domains. The corresponding magnetic phases are indicated by labels in c. The black frame in d highlights the area where additional anomalies appear in the multi-domain state which are magnetic states confined to polar domain walls. Adapted under terms of the CC-BY license.\textsuperscript{\cite{geirhos2020}} Copyright 2020, The Authors, published by Springer Nature.}
  \label{Fig_PFM}
\end{figure}

Beyond the macroscopic magnetic, electric, and optical properties of the rhombohedral lacunar spinels, also the local microscopic properties of ferroelectric domains and domain walls are fascinating \cite{neuber2018, geirhos2020,kezsmarki2015, butykai2017}. Especially, the growing interest in technologically relevant microscopic properties of domains and domain walls of oxide-based multiferroics \cite{meier2015, catalan2012, evans2020} may also trigger considerable interest in this novel non-oxide family of ferroelectrics. Through the examples of \GVS~\cite{butykai2017} and \GVSe~\cite{geirhos2020}, \textbf{Figure~\ref{Fig_PFM}}a and b reveal the typical polar domain patterns in ferroelectric lacunar spinels via scanning probe techniques.

These patterns are dominated by lamellar-like domain structures, formed by the alternation of two types of domains. Adjacent lamellar structures are separated by staircase-like domain walls \cite{neuber2018, geirhos2020,butykai2017}. A recent article \cite{geirhos2020}, reporting on the emergence of magnetic states confined to polar domain walls of \GVSe , implies that polar domain walls in multiferroics can offer a fertile ground to explore novel magnetic states non-existing in bulk materials. Through the electric control of the domain state (see section~\ref{pol}), comparative studies of the multi- versus mono-domain states in \GVSe\ are feasible.

Though the magnetic states confined to polar domain walls of \GVSe\ are present only in a small volume fraction, their signatures can be observed in macroscopic quantities. This was evidenced by anomalies in the magnetization and polarization exclusively observed in the multi-domain samples and assigned to spin textures emerging only at polar domain walls. Figure \ref{Fig_PFM}c and d, reproduced from ref.~\cite{geirhos2020}, show color maps of the dependence of the magnetocurrent on the magnitude and the orientation of the magnetic field, obtained on mono- and multi-domain samples, respectively.
While in the mono-domain case only anomalies associated with magnetic transitions of the $P_1$ domain are present (Figure~\ref{Fig_PFM}c), in the multi-domain state the same set of anomalies is observed twice, shifted by $\sim109^\circ$, the angle spanned by the rhombohedral axes of $P_1$- and $P_2$-type domains (Figure~\ref{Fig_PFM}d). Most importantly, additional anomalies show up in the multi-domain state (in the area marked by the black frame), which can not be assigned to any of the four polar domains and thus, must originate from magnetic states confined to ferroelectric domain walls \cite{geirhos2020}. 

These first intriguing results on the domain and domain-wall properties of rhombohedral lacunar spinels are already fuelling the idea of non-oxide based domain-wall electronics in skyrmion-hosting systems. The electric and magnetic field-control of macroscopic properties, directly linked to polar domain architectures, makes systems like \GVS, \GVSe, and \GMS\ prime candidates for exploratory research. Thorough local-probe studies are required and absolutely essential to explore dielectric, magnetoelectric, and piezoelectric properties of these compounds from the macro- to the nanoscale.

\medskip
\textbf{Acknowledgements} \par
Financial support by the Deutsche Forschungsgemeinschaft (project number 107745057, TRR 80) is gratefully acknowledged.

\medskip
\textbf{Conflict of Interest} \par
The authors declare no conflict of interest.

\medskip

%

\bibliographystyle{MSP}

\end{document}